\documentclass[fleqn,usenatbib]{mnras}
\usepackage[utf8]{inputenc}
\DeclareUnicodeCharacter{5415}{Lyu}
\DeclareUnicodeCharacter{6768}{yang}

\usepackage{newtxtext,newtxmath}

\usepackage[T1]{fontenc}

\DeclareRobustCommand{\VAN}[3]{#2}
\let\VANthebibliography\thebibliography
\def\thebibliography{\DeclareRobustCommand{\VAN}[3]{##3}\VANthebibliography}

\usepackage{graphicx}	
\usepackage{amsmath}	

\usepackage{subcaption}
\usepackage{longtable}

\title[X-ray AGN in the NewAthena era]{Unveiling faint X-ray AGN populations in the NewAthena era: Insights from cosmological simulations}

\author[N. Covas et al.]{
Nuno Covas,$^{1,2}$\thanks{E-mail: nunocovas27@gmail.com}
Israel Matute,$^{1,2}$
Stergios Amarantidis,$^{3}$
José Afonso,$^{1,2}$
Giorgio Lanzuisi,$^{4}$
\newauthor
Andrea Comastri,$^{4}$
Stefano Marchesi,$^{5,6,4}$
Ciro Pappalardo,$^{1,2}$
Rodrigo Carvajal,$^{1,2}$ and
Polychronis Papaderos$^{7}$
\\
$^{1}$Instituto de Astrofísica e Ciências do Espaço, Universidade de Lisboa- OAL, Tapada da Ajuda, PT1349-018 Lisboa, Portugal\\
$^{2}$Departamento de Física, Faculdade de Ciências da Universidade de Lisboa,Edifício C8, Campo Grande, PT1749-016 Lisboa, Portugal\\
$^{3}$Institut de Radioastronomie Millimétrique (IRAM), Avenida Divina Pastora 7, Local 20, E-18012, Granada, Spain\\
$^{4}$INAF, Osservatorio di Astrofisica e Scienza dello Spazio (OAS) di Bologna, via P. Gobetti 93/3, 40129 Bologna, Italy\\
$^{5}$Dipartimento di Fisica e Astronomia (DIFA), Università di Bologna, via Gobetti 93/2, I-40129 Bologna, Italy\\
$^{6}$Department of Physics and Astronomy, Clemson University, Kinard Lab of Physics, Clemson, SC 29634, USA\\
$^{7}$Instituto de Astrofísica e Ciências do Espaço, Universidade do Porto – CAUP, Rua das Estrelas, PT4150-762 Porto, Portugal\\
}

\date{Accepted XXX. Received YYY; in original form ZZZ}

\pubyear{\the\year{}}

\begin{document}
\label{firstpage}
\pagerange{\pageref{firstpage}--\pageref{lastpage}}
\maketitle

\begin{abstract}
Recent observations expanded our understanding of galaxy formation and evolution, yet key challenges persist in the X-ray regime, crucial for studying Active Galactic Nuclei (AGN). These limitations drive the development of next-generation observatories such as ESA’s NewAthena. Now in phase B (preliminary design), the mission requires extensive testing to ensure compliance with its scientific goals, particularly given the uncertainties surrounding high redshift AGN. This work leverages the IllustrisTNG cosmological simulation to build an X-ray AGN mock catalogue and assess the performance of NewAthena’s WFI. We created a Super Massive Black Hole (SMBH) light cone, spanning 10 $\mathrm{deg^2}$, with corrections to account for the limited resolution of the simulation and X-ray properties derived in post-processing. The resulting catalogue reveals a $\sim$5× overabundance of faint AGN compared to current X-ray constraints, an inconsistency potentially resolved by invoking a higher Compton-thick (CTK) fraction and intrinsic X-ray weakness, as suggested by recent JWST findings. An end-to-end survey simulation using SIXTE predicts $\sim$250,000 AGN detections, including $\sim$20,000 at z > 3 and 35 in the Epoch of Reionization (z > 6); notably, only AGN with $\mathrm{L_X} > 10^{43.5}$ erg $s^{-1}$ are detectable at $z > 6$. The analysis also forecasts a significant population of detectable CTK AGN, even beyond z > 4. These findings suggest X-ray observations will, for the first time, probe a significant AGN population in the EoR, offering new insights into SMBH growth. They also provide key input for refining NewAthena’s mission design and optimizing its survey strategy.
\end{abstract}




\section{Introduction} \label{Introduction}

One of the major goals of contemporary astrophysics is to unravel the complex processes governing galaxy formation and evolution. Among the most significant discoveries is the phenomenon of Active Galactic Nuclei (AGN), a luminous and compact source that can outshine the emission from the entire host galaxy. Currently, it is widely established that this emission arises from the accretion of matter into the Super Massive Black Hole (SMBH) present at the centre of every massive galaxy (e.g.,\citealt{kormendy_inward_1995, volonteri_formation_2010}).

Given the prevalence of SMBHs within galaxies, it is plausible that any presently inactive galaxy may have previously hosted an AGN. Furthermore, since the energy emitted by the AGN can exceed that emitted by its host, the coupling of this energy, through feedback mechanisms, can significantly impact the host (e.g., \citealt{fabian_observational_2012,morganti_many_2017}). Moreover, we have evidence of coevolution and self-regulation between the SMBH and the host, in the ubiquity of the scaling relations between the mass of the SMBH and galactic properties, such as the $M_{\bullet}-\sigma$ relation (\citealt{kormendy_coevolution_2013}; see also \citealt{donofrio_past_2021} for a recent review).

For these reasons, it is crucial to understand the growth of SMBHs as a function of redshift (e.g., \citealt{yang_linking_2018}) and the duty cycle of AGN (i.e. the fraction of time a galaxy hosts an AGN). Achieving this requires a detailed demographic analysis of AGN populations, beginning with a complete and systematic census of their properties.

\vspace{-1.25mm}

The X-ray regime is commonly favoured for AGN census (e.g., \citealt{brandt_cosmic_2015}), due to several advantages. Firstly, AGN X-ray emission typically surpasses that of stellar processes within the host, ensuring that X-ray surveys maintain a high purity in contrast to infrared surveys. Moreover, hard X-rays are relatively unaffected by obscuration, rendering X-ray surveys more complete and unbiased compared to optical surveys (e.g., \citealt{hickox_obscured_2018}). Finally, X-ray emission is a good tracer of accretion in AGN because it originates from the innermost parts of the accretion disk around the SMBH (e.g., \citealt{jovanovic_x-ray_2009}).

While current X-ray surveys have been indispensable to probe the properties of AGN, they have a few limitations. We currently lack a complete and unbiased knowledge of the faintest sources, such as high redshift (z>3) AGN \citep{pouliasis_active_2024} or the Compton Thick population (i.e. AGN with a line of sight column density, $\mathrm{N_H}$ > $\mathrm{1.5\times10^{24}}$ $\mathrm{cm^{-2}}$). Although current facilities such as XMM-Newton or Chandra have been used to place some constraints on these populations, the detailed characterization of the fainter AGN population will require telescopes with significantly better photon collecting capacity. 

These observational limitations highlight the need for next-generation X-ray observatories, as recognized in ESA's Cosmic Vision program "The Hot and Energetic Universe" \citep{nandra_hot_2013}, which led to the selection of ATHENA as the second Large-class Mission (L2). ATHENA combines a large effective area, high angular resolution, and a wide Field of View (FOV). Its original design featured a single grazing incidence telescope with a 12 m focal length \citep{willingale_hot_2013}, an effective area of 1.4 $\mathrm{m^2}$ at 1 keV, a 5" on-axis half energy width (HEW), a FOV exceeding 40'$\times$40', and a homogeneous Point Spread Function (PSF). This breakthrough is possible with the silicon pore optics (SPO) technology \citep{collon_development_2017,barriere_silicon_2022}, which ensures a high ratio of collecting area to mass while maintaining the required angular resolution. The mission incorporates two instruments: the Wide Field Imager (WFI; \citealt{antonelli2024wide}) and the X-ray Integral Field Unit (X-IFU; \citealt{barret_athena_2018,barret_athena_2023}). The WFI is a silicon detector using DEPFET Active Pixel Sensor (APS) technology, with a 40´×40´ FOV, a 0.2–12 keV spectral range, 80–170 eV energy resolution, and an 80 $\mu$s time resolution. The X-IFU is a cryogenic spectrometer with an unprecedented 2.5 eV energy resolution at 7 keV and a 5-arcmin FOV. The groundbreaking potential of ATHENA was demonstrated by comprehensive end-to-end simulations by several groups. \citet{marchesi_mock_2020} assessed the performance of ATHENA's WFI survey concerning high redshift AGN, while \citet{zhang_high-redshift_2020} conducted a similar study focusing on extended sources

In 2022, ATHENA’s development paused due to cost concerns. ESA later conceived NewAthena \citep{cruise_NewAthena_2024}, a rescoped concept retaining flagship status and most of ATHENA’s capabilities, targeting adoption by 2027 and a 2037 launch. NewAthena features a smaller mirror with an effective area of $\sim$1 $\mathrm{m^2}$ at 1 keV and a broader PSF with a better-than 9" on-axis HEW. The WFI largely maintains its original specifications, while the X-IFU will have a reduced FOV (4´×4´) and a 4 eV energy resolution at 7 keV. This reconfiguration led to a reduction in sensitivity and resolution, consequently, it is essential to revisit the end-to-end simulations to evaluate how these changes affect the mission's outcomes and maximize the scientific return within these constraints.

A fundamental ingredient of such simulations pertains the modelling of the input AGN, especially in the poorly constrained regions of the parameter space. Several approaches exist, such as extrapolations of the X-ray Luminosity Function (XLF); \citep{marchesi_mock_2020}. However, these methods are not physically motivated and have limited constraints at high redshift. Consequently, complementary methods should be considered.

Cosmological hydrodynamical simulations (HDS) are particularly insightful because they model galaxy formation by solving the equations of gravity and hydrodynamics in a cosmological context, tracing the properties of dark matter, gas, and stars in discrete resolution elements (see \citealt{somerville_physical_2015} and \citealp{vogelsberger_cosmological_2020} for reviews), albeit at a significant computational cost. Despite their strengths, they rely on subgrid prescriptions to represent unresolved processes, such as SMBH accretion and AGN feedback, that have been instrumental to the success of these models in reproducing observations (see \citealp{di_matteo_massive_2023} for an extensive review on the role of SMBHs in recent cosmological simulations). 

By simulating galaxy formation from physically motivated prescriptions, these models offer invaluable insights into AGN in the high redshift universe, where observations are limited, making them ideal for creating mock catalogues to test NewAthena. Moreover, they provide direct access to SMBH properties, such as masses, accretion rates, and AGN clustering. This framework has been extensively used in the literature, to provide predictions for surveys such as the XXL survey with XMM-Newton, \citep{koulouridis_xxl_2018} or eROSITA surveys, \citep{biffi_agn_2018,comparat_active_2019}. Previous work also ventured into making predictions for ATHENA, as evidenced by \citet{habouzit_supermassive_2021}, albeit their approach primarily involved analytical predictions (akin to \citealt{amarantidis_first_2019}), without delving into an end-to-end simulation of the survey.

In this work, we leverage a cosmological simulation of galaxy formation to create an X-ray mock catalogue of AGN, and simulate the X-ray sky to make quantitative predictions on the performance of NewAthena’s WFI survey. Our focus is on evaluating its capability to "Measure the space density of AGN that drive SMBH growth up to the Epoch of Reionization"\footnote{https://www.the-athena-x-ray-observatory.eu/en/unveiling-universe-x-rays}.

The paper is organized as follows: Sect. \ref{A Cosmological AGN Mock Catalogue} describes the methodology to create the mock catalogue and the observational comparisons used for validation. Section \ref{WFI Survey Simulation} presents SIXTE, the software used to simulate the mock observations, discusses the simulation strategy and outlines the detection process. In Sect. \ref{Results} we present the results that are discussed in Sect. \ref{Discussion}. We present our conclusions in Sec. \ref{Summary and conclusions}. We assume a flat $\mathrm{\Lambda CDM}$ cosmology with $\mathrm{H_0}$ = 67.4 $\mathrm{kms^{-1}Mpc^{-1}}$, $\mathrm{\Omega_m = 0.315}$ and $\mathrm{\Omega_{\Lambda}}$ = 0.685 \citep{planck_collaboration_planck_2020}.

\section{A Cosmological AGN Mock Catalogue}\label{A Cosmological AGN Mock Catalogue}

\subsection{IllustrisTNG} \label{IllustrisTNG}

In this work, we adopted the IllustrisTNG cosmological simulation \citep{nelson_illustristng_2021}, a state-of-the-art HDS. It consists of three flagship runs, in three different cosmological volumes: TNG50, TNG100, and TNG300, corresponding to box lengths of 50, 100, and 300 cMpc, respectively. We chose the TNG300 box, which is large enough to provide a statistically robust sample of AGN, essential for studying high redshift AGN. TNG300 has a volume of approximately $(300$ $\mathrm{cMpc})^3$, achieves a Dark Matter (DM) mass resolution of $\mathrm{M_{DM}}$ = 5.9$\times$$10^7$ $\mathrm{M_{\odot}}$, and a Baryonic Matter (BM) mass resolution of $\mathrm{M_{BM}}$ = 1.1$\times$$10^7$ $\mathrm{M_{\odot}}$.

IllustrisTNG builds upon the success of the original Illustris simulation \citep{nelson_illustris_2015}, incorporating significant improvements in the numerical implementation, and an updated physical model. The TNG model introduces magnetic fields, refines the model for the galactic winds, and updates the stellar evolution and chemical enrichment processes (see \citealt{pillepich_simulating_2018} for a detailed description and comparison with Illustris). The model also includes a new prescription for SMBH growth and dual mode AGN feedback \citep{weinberger_simulating_2017}. This new model alleviates some of the observational tensions in Illustris and produces more realistic galaxy populations (e.g., \citealt{naiman_first_2018, nelson_first_2018, springel_first_2018, marinacci_first_2018, pillepich_first_2018}).

The new SMBH model places seed SMBHs, with a mass of $1.2\times10^6$ $\mathrm{M_{\odot}}$, in dark matter halos with a mass greater than $7.4\times10^{10}$ $\mathrm{M_{\odot}}$, that do not contain an SMBH. Once seeded, these SMBHs grow through merging or Bondi-Hoyle accretion\footnote{Spherically symmetric accretion by a point source moving at a steady speed through a uniform gas cloud (\citealt{edgar_review_2004}).} given by

\begin{equation}
    \dot{M} = \frac{\alpha 4 \pi  (G M)^2 \rho}{(c_s^2 + v_{\infty}^2)^{3/2}},
    \label{eq:bondi_hoyle}
\end{equation}

where $\dot{M}$ is the accretion rate, $G$ is the gravitational constant, $M$ is the mass of the SMBH, $\rho$ is the local gas density, $c_s$ is the local sound speed, $\alpha$ is the boost factor that takes into account the unresolved Interstellar Medium (ISM) and $v_{\infty}$ is the gas speed around the SMBH. In the TNG model, $\alpha$ = 1, $v_{\infty}$ = 0 and the accretion is capped at the Eddington limit: 

\begin{equation}
    \dot{M}_{\text{Edd}} = \frac{4 \pi G M m_p}{\epsilon_r \sigma_T c},
    \label{eq:eddington_accretion}
\end{equation}

where $m_p$ is the proton mass, $\epsilon_r$ is the radiative efficiency, which IllustrisTNG adopts as 0.2, and $\sigma_T$ is the Thompson scattering cross section.

The model implements two modes of AGN feedback: thermal and kinetic modes. If the Eddington ratio ($\lambda$ = $\frac{\dot{M}}{\dot{M}_{\text{Edd}}}$) exceeds a value of $\min[0.002(\frac{M}{10^8M_\odot})^2, 0.1]$, thermal energy is injected into the neighbouring gas particles at a rate given by $\epsilon_{f,\text{high}} \epsilon_r \dot{M} c^2$. If the Eddington ratio is below this value, kinetic energy is injected into the gas at regular intervals of time, in the form of a wind oriented along a random direction. The energy injected is given by $\epsilon_{f,\text{low}} \dot{M}c^2$ and the coupling efficiencies $\epsilon_{f,\text{high}}$ and $\epsilon_{f,\text{low}}$ are free parameters of the model. For further details, the reader is referred to \citet{weinberger_simulating_2017} (see also \citealt{weinberger_supermassive_2018} and \citealt{terrazas_relationship_2020} for a detailed exploration of the impact of this improved feedback model).

\begin{figure}
   \centering
   \includegraphics[width=0.9\columnwidth]{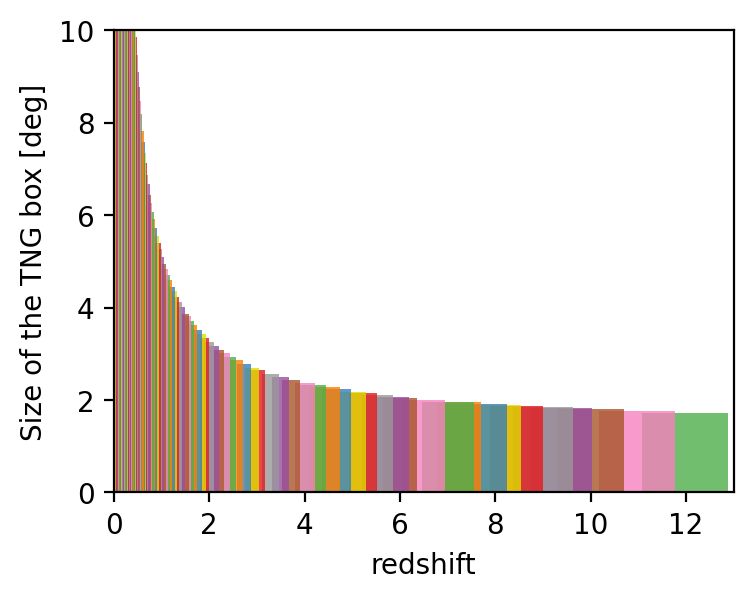}
      \caption{Graphical representation of the angular size as a function of redshift in TNG300. The x-axis shows snapshot redshifts, and the y-axis indicates the projected angular box size (300 cMpc) in degrees. Bar widths correspond to the redshift range of each box, with alternating colors for clarity. Due to the large number of snapshots, the redshift ranges of all boxes overlap significantly.}
     \label{fig:snapshots_figure}
   \end{figure}

Before attempting to create a mock catalogue from this simulation, it is insightful to verify whether there is enough data to cover the relevant redshift range. To do so, one of the axes of the box must be converted to a dimension in redshift, by inverting the following equation:

\begin{equation}
   d_c(z) = \frac{c}{H_0} \int_{z_1}^{z_2} \frac{dz'}{\sqrt{\Omega_m (1+z')^3 + \Omega_\Lambda}},
    \label{eq:comoving_distance}
\end{equation}

where $d_c(z)$ is the comoving distance in the redshift interval between $z_1$ and $z_2$, $H_0$ is the Hubble constant and $\Omega_m$,  $\Omega_\Lambda$ are the matter and dark energy density parameters. With this prescription, each of the 100 snapshots from TNG300 can be assigned to a specific redshift interval, centred on the fiducial redshift of the snapshot, as illustrated in Fig. \ref{fig:snapshots_figure}. 

\subsection{Building the light cone} \label{Building the light cone}

The raw output of the simulation are discrete snapshots of the simulated box, in the comoving rest frame (i.e. removing the effect of cosmological expansion). These cannot be directly used as input for the end-to-end simulations; instead, a continuous light cone must be created,  which we have implemented in the following way \citep{amarantidis_towards_2021}:

\begin{enumerate}

    \item For each snapshot, all SMBH particles are retrieved\footnote{No mass cut has been applied to the catalogue, therefore the newly-seeded SMBHs are included in the analysis. However, due to their low accretion rate, they don’t contribute significantly to the results and are mainly a source of unresolved background.}. Each particle carries information about its identification, position, BH mass, accretion rate, nearby gas density, the total mass of the particle (BH+gas reservoir) and the mass of the host halo.   
    
    \item Due to the redshift overlap in the boxes (see Fig. \ref{fig:snapshots_figure}), care must be taken to not duplicate the SMBH in overlapping regions. Therefore, for each snapshot, we only keep the sources in a limited slice. The width of this slice is calculated as the difference between the comoving distances at the redshift of the current snapshot and at the redshift of the following snapshot. The slice is then sampled from a random position in the box.
    
    \item Since the simulation evolves the same sources forward in time, their positions remain roughly consistent across snapshots. To avoid biasing the catalogue, each box is rotated ±90° around a random axis. Although these rotations helped mitigate bias, some circular artefacts persisted. To address this, a random translation of the box centre is also applied, effectively removing the remaining artefacts, while preserving the structure and correlations between the sources in the slice.

    \item Since the snapshots are discrete, some processing is necessary to assign a different redshift to each source, ensuring a continuous light cone. First, the redshift of the snapshot is used to calculate a comoving distance, using Eq. \ref{eq:comoving_distance}, then the radial coordinate of each particle is added to this value, and the resulting distances are converted into redshifts for each particle by inverting Eq. \ref{eq:comoving_distance}. Each of these redshifts is then used to calculate a scale factor\footnote{using the function kpc\_comoving\_per\_arcmin from the astropy.cosmology package}
    \footnote{https//docs.astropy.org/en/stable/cosmology/index.html} that converts the transverse coordinates, in ckpc, to the projected sky coordinates Right Ascension and Declination, in arcmin.
    
    \item The fixed dimension of the simulated box, 300 cMpc, limits the angular size of the light cone that can be constructed from it. The high redshift scale factor, at z = 12, limits the light cone to a projected angular size of 1.6 deg, as plotted in Fig. \ref{fig:snapshots_figure}, corresponding to an area of 2.5 $\mathrm{deg^2}$\footnote{The actual size is 1.7 deg, however, the light cone is further limited to ensure some margin, for the translations of the box centre.}. Considering this, 4 unique light cones were created from replications of the TNG300 box. Thus, the final light cone covers an area of simulated sky equivalent to 10 $\mathrm{deg^2}$ in a redhsift range from 0 to 12. 
    
\end{enumerate}

\subsubsection{Resolution correction} \label{Resolution correction}

In addition to the physical model, there are two critical factors in cosmological simulations: the volume of the box and the resolution of the particles. Large volumes are necessary to simulate a representative universe, with the rarest and most massive AGN; a high resolution is necessary to ensure accurate model results. There is always a tradeoff between these parameters; a larger volume of the simulated box necessitates a lower resolution to maintain computational feasibility. The IllustrisTNG simulations also follow this general trend: the TNG300-1 box has a volume 27 times larger than the volume of TNG100-1, however, the spatial and mass resolutions are lower by a factor of 2 and 8, respectively. 

As discussed in Appendix B of \citet{weinberger_supermassive_2018}, the resolution has a significant impact on the properties of galaxies and SMBH. The accretion rates and masses, as a function of the Host Halo mass, are systematically lower, compared to the high-resolution simulation. This is especially important for SMBH in lower mass Halos (< $10^{12}$ $\mathrm{M_{\odot}}$) since the Halos are resolved by few particles and the SMBH are not in the self-regulating regime (kinetic feedback mode). This greatly impacts the bright end of the XLF, and it is critical for high redshift predictions, where most of the SMBH exists in these Halos. 

To study model convergence, the TNG project repeated all flagship runs at a lower resolution; TNG100-2 has the same volume and initial conditions as TNG100-1, but the same resolution as TNG300-1. Comparing these simulations allows one to account for the impact of the lower resolution. This has been done in \citet{pillepich_first_2018}, where the authors determined simple corrections for the stellar masses and star formation rates of TNG300-1. Motivated by that work, we derived corrections for the distributions of the Eddington ratio of the SMBH. The details are in Appendix \ref{Resolution Correction for the Eddington ratio}.

\begin{figure}
\centering
\includegraphics[width=\columnwidth, height = 45mm]{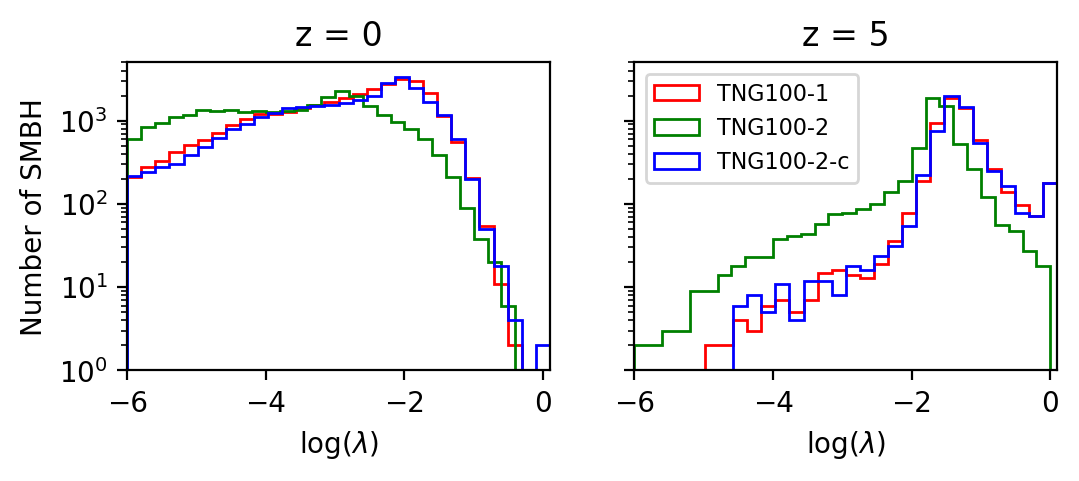}
\caption{Histograms of log($\lambda$) distributions at z = 0 (left) and z = 5 (right) for TNG100-1 (red), TNG100-2 (green), and the corrected TNG100-2-c (blue). The y-axis (logarithmic scale) shows frequencies, highlighting the similarity between TNG100-1 and TNG100-2-c at both redshifts.}
\label{fig:Edd_rate}
\end{figure}

The results of these corrections are illustrated by Fig. \ref{fig:Edd_rate}, which plots the histograms of the Eddington ratio distributions of TNG100-1 and TNG100-2, as well as the corrected distribution, TNG100-2c. The figure shows that the high and low resolution simulations produce different distributions (Fig \ref{fig:Edd_rate}, red and green histograms respectively) and the corrections are successful as the corrected distribution of Eddington ratios (blue histogram) is similar to the high-resolution simulation (red histogram). The K-S test has a p-value of 0.27 and 0.79 at z = 0 and z = 5, respectively, therefore the TNG100-1 and TNG100-2-c distributions are statistically indistinguishable. Comparing TNG100-2 and TNG100-2-c, at z = 0, the high-lambda end of the distribution is similar; however, this is not the case at z = 5, where the correction considerably increases the number of SMBH accreting at the Eddington limit. This has important implications regarding the X-ray luminosity of these AGN and consequently, their detectability with future X-ray observatories such as NewAthena.

\subsection{X-ray properties} \label{X-ray properties}

IllustrisTNG does not provide X-ray properties for the AGN; therefore, these are determined in post-processing, following several models and empirical assumptions.

\subsubsection{X-ray luminosity} \label{X-ray luminosity}

The model adopted to calculate the bolometric luminosity assumes two distinct modes of accretion: a Shakura-Sunyaev thin disk for high Eddington ratio AGN \citep{shakura_black_1973}, and an Advection Dominated Accretion Flow (ADAF) for low Eddington ratio AGN (e.g., \citealt{yuan_hot_2014}). For ADAF, the radiative efficiency depends on the Eddington ratio \citep{mahadevan_scaling_1997,churazov_supermassive_2005,hirschmann_cosmological_2014}, resulting in a decrease of this value compared to the thin disk model. The reader is referred to \citet{yuan_hot_2014} for a review on ADAF and other hot accretion flows. This model can be parameterized in the following way: 

\begin{equation}
L_{Bol} =
    \begin{cases}
        \epsilon_r \dot M c^2 & \text{if } \lambda >= 0.1\\
        \frac{0.2}{\alpha^2}\lambda \epsilon_r \dot M c^2 & \text{if } \lambda < 0.1\\
    \end{cases}
\label{L_Bol_model}
\end{equation}

where $\mathrm{\dot M}$ is the accretion rate, $\mathrm{\dot M_{Edd}}$ is the accretion rate at the Eddington limit, $\lambda$ = $\frac{\dot M}{\dot M_{Edd}}$ is the Eddington ratio, $\epsilon_r$ = 0.08 is the radiative efficiency we assume (see Sec. \ref{Mock catalogue validation}), c is the speed of light, and $\alpha$ is the viscosity parameter.

This model is very similar to the one adopted in previous work with IllustrisTNG
\citep{weinberger_supermassive_2018,habouzit_linking_2019,habouzit_supermassive_2021}. The main difference is the numerical prefactor of the ADAF.  Previous works assumed a constant factor of 10, to ensure continuity between the branches of the function; however, as explained by \citet{griffin_evolution_2019}, this prefactor depends on the radius of the Innermost Stable Circular Orbit (ISCO), which depends on the spin of the SMBH. Since IllustrisTNG does not track this quantity, we model the ADAF following Eq. 58 of \citet{mahadevan_scaling_1997}, which assumes canonical values for the unconstrained parameters, and adopt $\alpha$ = 0.3, consistent with the findings of \citet{liu_constraints_2013}. 

After the bolometric luminosities were calculated, we determined the X-ray luminosities assuming a bolometric correction, $K_X = \frac{L_{Bol}}{L_X}$. We adopt the hard X-ray (2-10 keV) corrections from \citet{duras_universal_2020}, since these are the most up-to-date bolometric corrections for X-ray, and are valid in a range of 7 dex in bolometric luminosity:

\begin{equation}
K_X(\lambda) = 7.51\left(1+\left(\frac{\lambda}{0.05}\right)^{0.61}\right)
\label{Bolometric_Corrections}
\end{equation}

To model the scatter, due to the variability, which is not resolved by these simulations \citep{habouzit_supermassive_2021}, as well as the intrinsic X-ray weakness in some AGN populations \citep{pu_fraction_2020,laurenti_x-ray_2022, maiolino_jwst_2024, lyu2024active}, the value of the bolometric correction is randomly sampled from a Gumbel distribution, centred on the relation from \citet{duras_universal_2020}, with a width of 0.5 dex \citep{habouzit_supermassive_2021}. 

Finally, these X-ray luminosities are converted into X-ray fluxes, using the formula: 

\begin{equation}
F_{X} = \frac{L_{X}}{4\pi(d(1+z))^2K(z)},
\label{flux}
\end{equation}

where $\mathrm{L_X}$ is the 2-10 keV X-ray luminosity, $z$ is the redshift of the AGN, d is the comoving distance at that redshift, and K($z$) is the K-correction. The K-corrections are calculated using the spectral templates (see Sec. \ref{AGN spectral templates}) assumed for each AGN.

\subsubsection{Obscuration model} \label{Obscuration model}

X-ray AGN are typically obscured by gas within the line-of-sight, due to the photoelectric absorption, at energies below 10 keV, or Compton scattering at higher energies. Therefore, we must assume an obscuring column density, $\mathrm{N_H}$, for each AGN \citep{hickox_obscured_2018}. The model we adopted is similar to the radiation-lifted torus model developed by \citet{buchner_galaxy_2017} and \citet{buchner_galaxy_2017-1}. It comprises three components: CTK obscuration, nuclear Compton Thin obscuration (CTN), and CTN obscuration from the host. 

Regarding CTK obscuration, despite substantial efforts by multiple teams to directly constrain it through X-ray surveys (e.g., \citealt{lanzuisi_chandra_2018, georgantopoulos_nustar_2019, lambrides_large_2020, torres-alba_compton-thick_2021, laloux_demographics_2022, yan_most_2023, carroll_high_2023}) significant uncertainties and discrepancies persist. A complementary approach to quantifying this heavily obscured AGN population involves population synthesis models that fit the Cosmic X-ray Background (CXB) (e.g.,\citealt{gilli_synthesis_2007,ueda_toward_2014}) and estimate the CTK fraction. The most recent model of this kind, developed by \citet{ananna_accretion_2019}, provides improved consistency with X-ray observational constraints and predicts an average CTK fraction of 50\% in the local universe.

However, the most recent observations with the James Webb Space Telescope (JWST) find many AGN (candidates) that are remarkably X-ray weak \citep{maiolino_jwst_2024,lyu2024active} and one of the leading explanations for this phenomenon is a significant increase in the CTK fraction with redshift, for intermediate luminosity ($L_{Bol}$ < $10^{11}$ $L_\odot$) AGN. Considering this, our model assumes a CTK fraction parameterized as:

\begin{equation}
    F_{CTK} = 0.4 + 0.5\frac{1-e^{-z}}{1+e^{logL_X-44}},
\end{equation}

with $\mathrm{N_H}$ drawn randomly from a log-uniform distribution, between $10^{24}$ $\mathrm{cm^{-2}}$ and $10^{26}$ $\mathrm{cm^{-2}}$. For further discussion on these assumptions, the reader is referred to Sec. \ref{Mock catalogue validation}

Unlike the CTK obscuration, the CTN obscuration is much better constrained, with several works finding an evolution with redshift and an anti-correlation with the luminosity (e.g., \citealt{ueda_toward_2014,peca_cosmic_2023}), therefore, most obscuration models assume an explicit luminosity dependence for this fraction. However, the work by the BASS team on the radiation-regulated growth model for AGN \citep{ricci_close_2017,ricci_bass_2022,ananna_probing_2022,ricci_bass_2023} shows that, in the local universe, the main driver of nuclear obscuration is the Eddington ratio. For this reason, the nuclear CTN fraction is calculated based on a sigmoid fit to the X-ray covering fraction in Fig. 3 of \cite{ricci_bass_2023}:

\begin{equation}
F_{CTN} = 0.6\frac{10^{-1.21(x+1.525)}}{1+10^{-1.21(x+1.525)}}+0.25,
\label{covering_fraction}
\end{equation}

where x = $log_{10}(\lambda)$, with $\lambda$ the Eddington ratio and $N_H$ is chosen randomly from a log-uniform distribution, between $10^{22}$ $\mathrm{cm^{-2}}$ and $10^{24}$ $\mathrm{cm^{-2}}$. Although this relation was derived using local AGN, we assume its validity for all $z$, (see \citealt{jun_dust--gas_2021} for a discussion on the $N_H-\lambda$ relation at higher $z$).

The last component of the model is the obscuration by the host. Although this component might not be very important at low $z$, since it cannot exceed a column density of $10^{22}$ $\mathrm{cm^{-2}}$, recent works (e.g., \citealt{circosta_x-ray_2019,damato_dust_2020,gilli_supermassive_2022,signorini_x-ray_2023}) show that this might be the dominant source of obscuration at high $z$, even reaching CTK column densities. IllustrisTNG is a HDS, therefore, the obscuration from the host could be measured by ray-tracing of the gas particles, the approach followed by \citet{buchner_galaxy_2017-1}; however, that is beyond the scope of this work and instead a simplified model, based on the work by \citet{gilli_supermassive_2022} is used.

The obscuration from the host galaxy is modeled using a log-normal distribution with an average column density of $10^{21}(1+z)^{3.3}$ $\mathrm{cm^{-2}}$ and a standard deviation of 0.5 dex. In addition, a constant host covering fraction of 40\% is assumed. Given the significant uncertainty in the redshift evolution of the host covering fraction, we adopted the average value of the sample by \citet{buchner_galaxy_2017-1}.

After using the model to determine the column density for each AGN, the values are binned and used, with the spectral templates (see Sec. \ref{AGN spectral templates}), to calculate the observed flux in the soft (0.5-2 keV) and hard (2-10 keV) bands.

\subsubsection{AGN spectral templates} \label{AGN spectral templates}

\begin{figure}
\centering
\includegraphics[width=0.95\columnwidth]{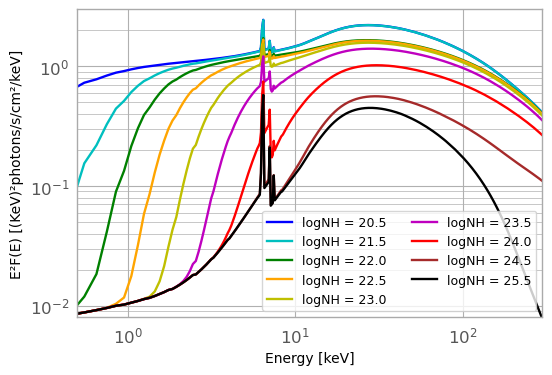}
\caption[Example spectral templates]{Example spectral templates employed for AGN in the mock catalogue, at z=0, delineated across various logNH values as depicted in the legend. All spectra have $\Gamma$ = 1.9.}
\label{fig:Spectra_plot}  
\end{figure}

To model X-ray AGN, some spectral shape must be assumed for each source. The templates used in this work, shown in Fig. \ref{fig:Spectra_plot}, were created using PyXspec \citep{gordon_pyxspec_2021}, a Python interface for the XSPEC software \citep{arnaud1996xspec}.

In XSPEC nomenclature, 3 different types of models are employed, according to three regimes of obscuration:

\begingroup
\setlength{\jot}{1pt}
\begin{align}
    &phabs*\big(\text{c}*\text{zcutoffpl} + 
    \text{zphabs}*\text{cabs}*(\text{pexmon})\big), \notag\\ 
    &\hspace{1em} \log N_H < 22 \label{eq:unobscured} \\
    &phabs*\big(\text{c}*\text{zcutoffpl} + 
    \text{zphabs}*\text{cabs}*(\text{zcutoffpl}) + 
    \text{pexmon}\big), \notag\\ 
    &\hspace{1em} 22 \leq \log N_H < 25 \label{eq:obscured} \\
    &phabs*\big(\text{c}*\text{zcutoffpl} + 
    \text{pexmon}\big), \notag\\ 
    &\hspace{1em} \log N_H \geq 25 \label{eq:very_obscured}
\end{align}
\endgroup

These spectral templates model key features commonly observed in X-ray spectra of AGN, such as a primary X-ray continuum, a reflection component, and an unabsorbed scattered component (\citealt{ueda_toward_2014, ricci_bat_2017, ananna_accretion_2019}).

The primary X-ray continuum is the result of the Comptonization of the thermal photons from the accretion disk, occurring in the hot corona above the disk. More specifically, the thermal photons undergo Inverse Compton scattering from the energetic electrons present in the corona, which increases their energy and prompts a cascade of more energetic scattering events. This results in a power-law spectrum, with a slope given by the photon index, $\Gamma$, and an exponential cutoff at high energy, since eventually the photons reach the energy of the electrons and the energy transfer becomes inefficient \citep{done_observational_2010}. 

Therefore, the primary continuum emission is modeled as a cutoff power law, with a cutoff energy, $\mathrm{E_c}$ = 200 keV, close to the mean value found in \citet{ricci_bat_2017}, with a Kaplan–Meier estimator; this value is also supported by NuSTAR observations (e.g., \citealt{akylas_distribution_2021}). Photoelectric absorption and Compton scattering are taken into account by the $zphabs$ and $cabs$ models, respectively. Both models assume the same $\mathrm{N_H}$ value.

This primary emission can then be transmitted, reflected, or scattered. The reflected component is modeled by $Pexmon$ \citep{nandra_xmmnewton_2007}, which assumes that the primary emission is reflected by the accretion disk. This model also takes into account self-consistently generated fluorescence lines, namely the Fe $\mathrm{K_{\alpha}}$, Fe $\mathrm{K_{\beta}}$, Ni $\mathrm{K_{\alpha}}$ and Fe $\mathrm{K_{\alpha}}$ Compton shoulder. $Pexmon$ assumes a cutoff power law, with the same shape and magnitude as the primary emission, the inclination angle of the accretion disk is fixed at 30º, while the reflection strength is R = 0.83, for the unobscured AGN, and R = -0.37 for the obscured AGN. These are the mean values of the reflection strength of \citet{ricci_bat_2017} when the upper and lower limits of all AGN are taken into account. These measurements are also consistent within 1$\sigma$ with the values of \citet{akylas_distribution_2021} for unobscured AGN, with NuSTAR.

It is important to note that in our assumption, all obscuration with a column density below $\mathrm{logN_H}$ < 22 is attributed to the host. Consequently, in Eq. \ref{eq:unobscured}, obscuration is accounted for in the reflected component. Conversely, in Eq.\ref{eq:obscured}, the reflected component remains unobscured. This approach is consistent with the templates presented in \citet{marchesi_mock_2020}, \citet{ricci_bat_2017} and \citet{ueda_toward_2014}.

The scattered component is modeled by a cutoff power law with the same shape and magnitude as the primary continuum, with a scattered fraction of 1\%, the mean value of \citet{ricci_bat_2017}. Although a recent investigation by \citet{gupta_bat_2021} identified a correlation between the scattered fraction and the column density among obscured sources, we opted for a simplified approach, assuming the mean value of their Swift-BAT sample in all spectral templates.

The $phabs$ model takes into account the local obscuration from the Milky Way, with a $N_H$ = $1.8\times10^{20}$ $\mathrm{cm^{-2}}$\citep{marchesi_mock_2020}. All other parameters, not explicitly mentioned, were left to their default values. The photon indices were randomly chosen from a normal distribution with a mean value of $\Gamma$ = 1.9 and a standard deviation of $\Delta \Gamma$ = 0.15 \citep{gilli_synthesis_2007}. These spectral templates cover a redshift range between 0 and 10, with a step of $\Delta z$ = 0.1; a photon index range between 1.7 and 2.1, with a step of $\Delta \Gamma$ = 0.1 and a $\mathrm{logN_H}$ range between 20.5 and 25.5, with a step of 0.5. Following \citet{marchesi_mock_2020}, the spectra with $\mathrm{logN_H}$ > 25 are modeled as reflection-dominated spectra, without a transmitted component.
The mock catalogue has a flux limit of $10^{-30}$ erg $s^{-1} cm^{-2}$.

\subsection{Mock catalogue validation} \label{Mock catalogue validation}

We tested the mock catalogue against avaialable constraints, namely the obscured fractions, the cumulative number counts, and the CXB flux. 

\begin{figure*}
\centering
\includegraphics[width=0.65\textwidth]{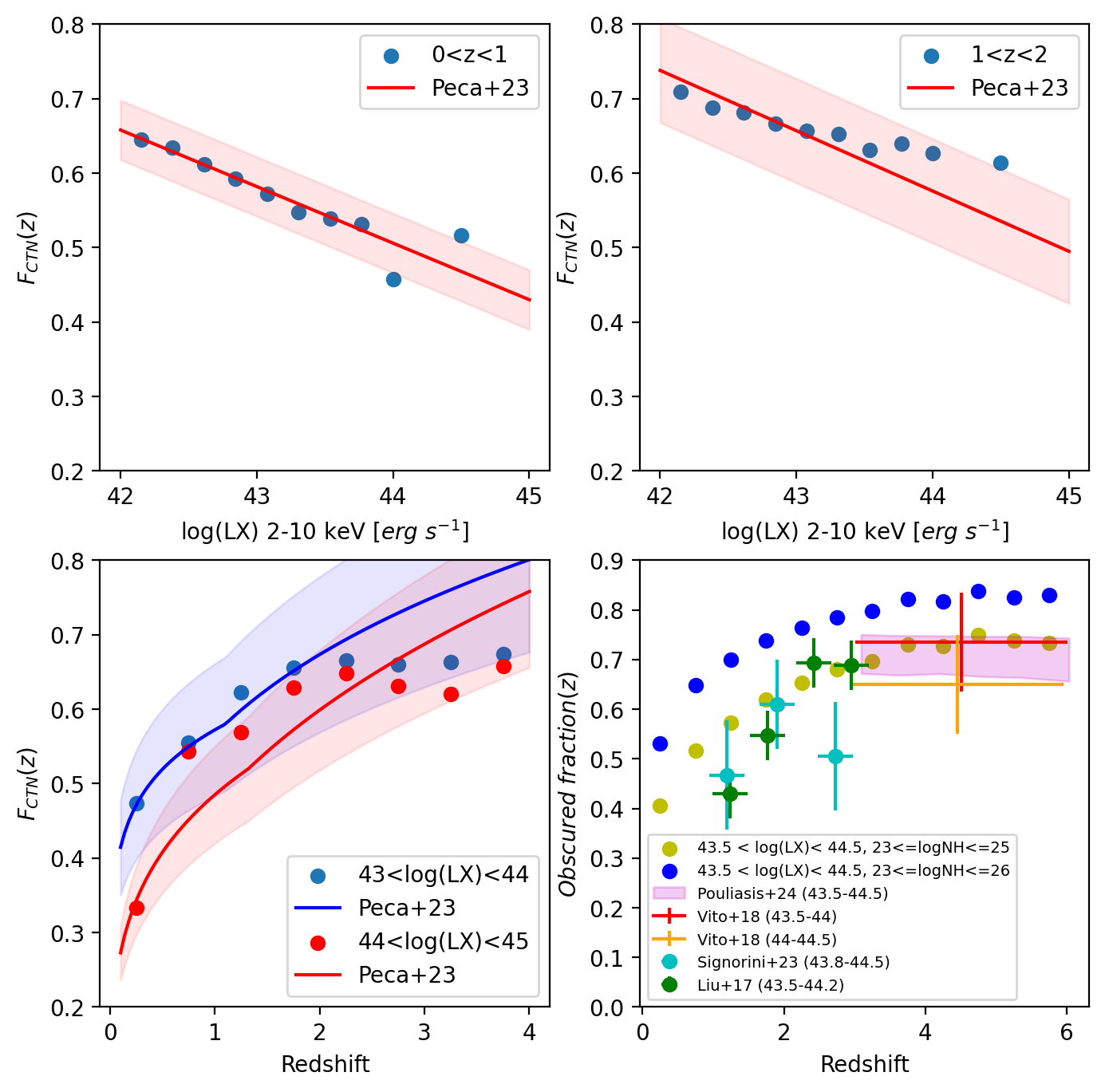}
\caption[Obscuration fraction plot]{
Top left panel: The Compton Thin fraction, $F_{CTN}(z)$, defined in Equation \ref{CTN}, in a redshift range $0<z<1$, is presented (blue dots) and compared to the recent work by \citet{peca_cosmic_2023} (red line fit and uncertainty band). Top right panel: The same as the top left panel, but for a redshift range $1<z<2$. Bottom left panel: The evolution of the $F_{CTN}(z)$ with redshift, for two luminosity ranges: $43<\log(L_X)<44$ (blue points) and $44<\log(L_X)<45$ (red points). The data is also compared with fits and uncertainty bands from \citet{peca_cosmic_2023}. Bottom right panel: The obscured fraction, now defined by Equation \ref{Obscured_fraction} (blue dots) or with $\log(N_H)<25$ (yellow dots), for a luminosity range of $43.5<\log(L_X)<44.5$, is presented as a function of redshift and compared with several works in the literature, namely: \citet{pouliasis_active_2024}, \citet{vito_high-redshift_2018}, \citet{signorini_x-ray_2023}, and \citet{liu_x-ray_2017}.}
\label{fig:Obscuration}
\end{figure*}

Figure \ref{fig:Obscuration} compares the obscured fractions in the mock catalogue with several measurements from the literature. The upper panels plot the CTN fraction, defined as:

\begin{equation}
    F_{CTN}(z) = \frac{N(22<=log(N_H)<=24)}{N(log(NH)<=24)}, 
    \label{CTN}
\end{equation}

as a function of the hard X-ray luminosity (2-10 keV), for redshift ranges 0<z<1 (left plot) and 1<z<2 (right plot) and compare the mock catalogue with the measurements from \citet{peca_cosmic_2023}. Despite not assuming a direct dependence on the luminosity, the adopted model successfully reproduces the observed anti-correlation at low z. The bottom left panel plots the CTN fraction as a function of $z$, for two luminosity ranges, 43<$\mathrm{log(L_X)}$<44 (blue) and 44<$\mathrm{log(L_X)}$<45 (red), again the mock catalogue is compared with the measurements from \citet{peca_cosmic_2023}. There is good agreement in the results for z<2, however, for higher z, the CTN fraction flattens at 65\%, while the best fits by \citet{peca_cosmic_2023} continue to increase. It should be noted, however, that this flattening is consistent with previous works (see Fig. 13 in \citealt{peca_cosmic_2023}; see also \citealt{aird_evolution_2015,buchner_obscuration-dependent_2015,ananna_accretion_2019}) and the fits are very uncertain due to the low number of high z AGN in this shallow survey.

The bottom right panel of Fig. \ref{fig:Obscuration} plots the obscured fraction, defined as 
\begin{equation}
   Obscured fraction (z) = \frac{N(23<=log(N_H)<=26)}{N(log(NH)<=26)}, 
    \label{Obscured_fraction}
\end{equation}

for a redshift range 0<z<6 and compares it with several measurements from the literature. This fraction seems to overestimate the constraints, however, this is not unexpected as current surveys are strongly biased against AGN with $N_H$ > $10^{25}$ $cm^{-2}$. If these AGN are excluded and the obscured fraction redefined, the agreement with the observations improves remarkably, further validating our assumptions regarding the CTK fraction. Overall, the obscuration model seems to comply with the observations.

\begin{figure*}
\centering
\begin{minipage}{0.48\textwidth}
    \centering
    \includegraphics[width=0.9\textwidth]{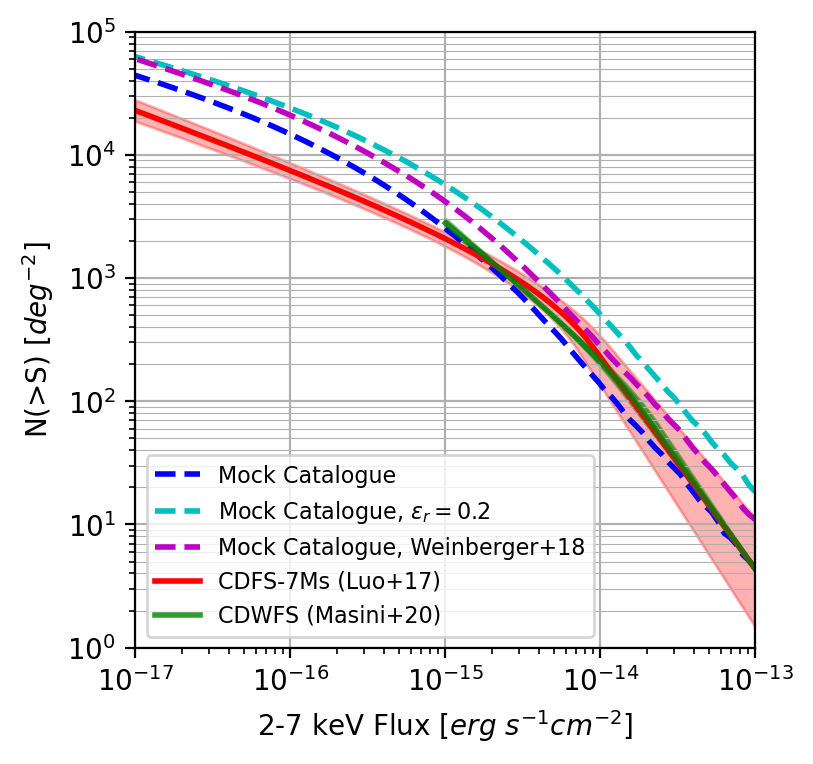}
\end{minipage}%
\hspace{0.01\textwidth}%
\begin{minipage}{0.48\textwidth}
    \centering
    \includegraphics[width=0.9\textwidth]{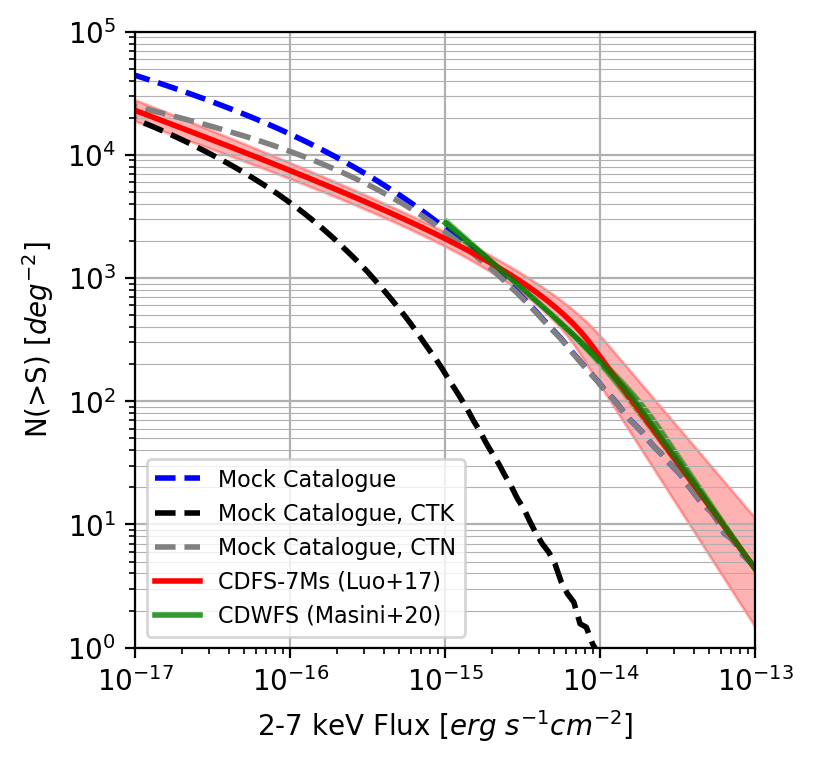}
\end{minipage}
\caption[Cumulative number counts comparison]{Comparison of cumulative AGN number counts as a function of 2-7 keV flux ($\mathrm{erg}$ $\mathrm{s^{-1} cm^{-2}}$) from the mock catalogue (blue) with observational datasets. Flux is derived from 2-10 keV flux using a power-law model ($\Gamma = 1.4$). The red line shows the CDFS-7Ms survey fit \citep{luo_chandra_2016}, and the green line represents CDWFS data \citep{masini_chandra_2020}, with corresponding confidence intervals (red/green bands). Predictions using $\epsilon_r$ = 0.2 (cyan) and the Bolometric Luminosity model \citep{weinberger_supermassive_2018} (magenta) are included. The second panel partitions counts into CTK and CTN categories.}
\label{fig:number_counts_side_by_side}
\end{figure*}

The next step in the validation process is to verify the cumulative number counts in the mock catalogue. Figure \ref{fig:number_counts_side_by_side} plots the cumulative number counts in the hard X-ray band (2-7 keV) and compares them with the measurements in Chandra Deep Field South-7Ms (CDFS-7Ms; \citealt{luo_chandra_2016}) and Chandra Deep Wide Field South (CDWFS; \citealt{masini_chandra_2020}). In the left plot of Fig. \ref{fig:number_counts_side_by_side}, the number counts show reasonable agreement above $10^{-15}$ $\mathrm{erg}$ $\mathrm{s^{-1} cm^{-2}}$. Below this flux the faint-end slope differs, resulting in an excess of faint AGN by a factor of $\sim$2, compared to CDFS-7Ms. These number counts are corrected for the incompleteness of the survey, which critically depends on the obscuration distribution of the AGN, that is not adequately constrained by \citet{luo_chandra_2016} (see \citet{lambrides_large_2020} for further details). Since their selection is biased against CTK AGN, this could explain the disagreement at the faintest fluxes.

Inspecting now the constraints of the CXB, the total flux in the hard band (2-10 keV) is $2.16\times10^{-11}$ $\mathrm{erg}$ $\mathrm{s^{-1} cm^{-2} deg^{-2}}$, while the total flux in the soft band (0.5-2 keV) is $0.88\times10^{-11}$ $\mathrm{erg}$ $\mathrm{s^{-1} cm^{-2} deg^{-2}}$. \citet{cappelluti_chandra_2017} measure the CXB spectrum and find total extragalactic fluxes of (2.034±0.005)$\times10^{-11}$ $\mathrm{erg}$ $\mathrm{s^{-1} cm^{-2} deg^{-2}}$ and (0.76±0.01)$\times10^{-11}$ $\mathrm{erg}$ $\mathrm{s^{-1} cm^{-2} deg^{-2}}$ in the hard (2-10 keV) and soft (0.5-2 keV) bands, respectively. The mock catalogue and this measurement differ by 6\% and 16\% in the hard and soft bands, respectively, which is within the calibration uncertainties, that could reach 20\% \citep{marchesi_mock_2020}. Given their uncertainties, the flux values from our mock catalogue agree with these measurements.

\citet{cappelluti_chandra_2017} also present results for the unresolved CXB, after removing all sources detected by \cite{civano_chandra_2016}, measuring fluxes of (6.47±0.82)$\times10^{-12}$ $\mathrm{erg}$ $\mathrm{s^{-1} cm^{-2} deg^{-2}}$ and (2.90±0.16)$\times10^{-12}$ $\mathrm{erg}$ $\mathrm{s^{-1} cm^{-2} deg^{-2}}$ in the hard and soft bands, respectively. Excluding all sources with a hard band (2-10 keV) flux greater than $3\times10^{-15}$ $\mathrm{erg}$ $\mathrm{s^{-1} cm^{-2} deg^{-2}}$ from the mock catalogue, (we assume that above this threshold the data from \citealt{civano_chandra_2016} is nearly complete), the remaining fluxes are $10.02\times10^{-12}$ $\mathrm{erg}$ $\mathrm{s^{-1} cm^{-2} deg^{-2}}$ and $3.46\times10^{-12}$ $\mathrm{erg}$ $\mathrm{s^{-1} cm^{-2} deg^{-2}}$, which differ by 55\% (2.75 $\sigma$) and 19\% ($\sim$1-sigma) \footnote{Assuming, again, that the uncertainty in this measurement is 20\%.} in the hard and soft bands, respectively.

This indicates that the flux of faint sources in the mock catalogue is in slight tension with the CXB, in the hard band\footnote{If a lower threshold of $2\times10^{-15}$ $\mathrm{erg}$ $\mathrm{s^{-1} cm^{-2} deg^{-2}}$ is adopted, the unresolved fluxes are $8.55\times10^{-12}$ and $2.77\times10^{-12}$ $\mathrm{erg}$ $\mathrm{s^{-1} cm^{-2} deg^{-2}}$ in the hard and soft band respectively, therefore, while the flux values are highly sensitive to the chosen threshold, the excess of flux in the hard band remains evident.}. In the right panel of Fig. \ref{fig:number_counts_side_by_side}, the number counts are partitioned into CTK and CTN number counts. Although the CTN number counts agree well with the observations, the inclusion of the CTK AGN leads to an overestimation. This suggests that IllustrisTNG overestimates the intermediate luminosity AGN population (log($L_X$)=42-44 erg $s^{-1}$), a conclusion that aligns with the findings of previous studies \citep{weinberger_supermassive_2018,habouzit_linking_2019, habouzit_supermassive_2021}.

Conversely, the issue could be in the modelling of the X-ray properties. It has been suggested that some AGN might be intrinsically X-ray weak \citep{pu_fraction_2020}, and this fraction might be significant for some populations (e.g., high Eddington ratio AGN; \citealt{laurenti_x-ray_2022}, see also \citealt{nardini_most_2019}). As discussed in the previous section, the mock catalogue accounts for this effect by assuming a distribution on the bolometric corrections, and the fraction of X-ray weak\footnote{X-ray weak by a factor of 10, relative to the relation from \citet{duras_universal_2020}} sources is 12.5\%. Consequently, assuming that IllustrisTNG provides an accurate representation of the universe, to comply with the CXB, both a very high CTK fraction and intrinsic X-ray weakness (with a fraction greater than 12.5\%) are required among the general AGN population.

The right panel of Fig. \ref{fig:number_counts_side_by_side} also illustrates the number counts, assuming a radiative efficiency of $\epsilon_r$ = 0.2 in the luminosity model (cyan line), the fiducial value assumed by IllustrisTNG. These number counts significantly exceed observational constraints, across the full flux range, indicating that a lower value is necessary to match the observed values. As \citet{sijacki_illustris_2015} points out for the original Illustris, in these models, the radiative efficiency is degenerate with the SMBH feedback efficiency, therefore, IllustrisTNG cannot uniquely constrain this parameter. For this reason, a value of $\epsilon_r$ = 0.08 was adopted, to ensure a reasonable match to the number counts. Fig. \ref{fig:number_counts_side_by_side} also presents the number counts based on the luminosity model by \citet{weinberger_supermassive_2018} (magenta line), assuming $\epsilon_r = 0.08$. These number counts also significantly exceed observational constraints, further illustrating the impact of the numerical prefactor in the low Eddington ratio branch of the model.

It is worth highlighting that while we assume a constant radiative efficiency and a constant prefactor on the ADAF, both of these quantities depend on the spin of the SMBH (\citealt{griffin_evolution_2019}; see also the discussion in \citealt{amarantidis_first_2019}), which is not tracked by IllustrisTNG. Considering the impact of these parameters on the number counts, it would be worthwhile to include subgrid modelling of the SMBH spin in future large-scale cosmological simulations. An example of subgrid spin modelling was already adapted to IllustrisTNG (in a small cosmological volume), as presented by \citet{bustamante_spin_2019}, 
and has shown success in mitigating some of the discrepancies observed in the fiducial run.

\subsection{Comparison with high redshift observations}  \label{Comparison with high redshift observations}

\begin{figure}
\centering
\includegraphics[width=\columnwidth, height = 55mm]{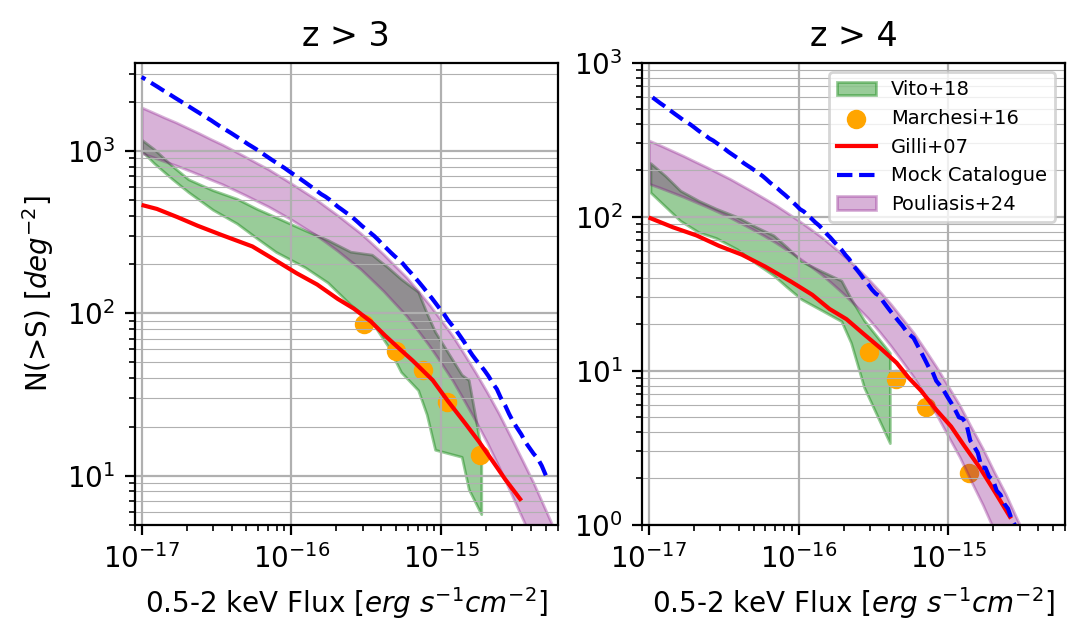}
\caption[High z number counts plot]{Comparison of cumulative high-redshift AGN number counts in the 0.5-2 keV flux band from the mock catalogue with observations. The left panel shows counts for $z>3$, and the right for $z>4$. Observational data are represented by the green-shaded region \citep{vito_high-redshift_2018} and orange points \citep{marchesi_chandra_2016}. Predictions from the mock catalogue (blue line), \citep{gilli_synthesis_2007} CXB model (red line) and the \citep{pouliasis_active_2024} XLF (purple band) are also shown.}
\label{fig:number_counts_high_z}
\end{figure}

We also plot the cumulative number counts at high z in Fig. \ref{fig:number_counts_high_z}, for z>3 (left plot) and z>4 (right plot), comparing with observations from \citet{marchesi_chandra_2016}, \citet{vito_high-redshift_2018} as well as the \citet{gilli_synthesis_2007} CXB model. Although these observations are consistent for higher fluxes, discrepancies arise at fainter flux levels, below $10^{-16}$ $\mathrm{erg}$ $\mathrm{s^{-1} cm^{-2}}$, highlighting the current uncertainty surrounding the high redshift AGN population. The mock catalogue exhibits a $\sim$3-fold excess of AGN at the faintest fluxes, highlighting a divergence with these observations.

The recent work by \cite{pouliasis_active_2024} studied the high redshift XLF and found a higher space density of AGN at z>3, compared to \cite{gilli_synthesis_2007}, \cite{marchesi_chandra_2016} and \cite{vito_high-redshift_2018}, most significantly for $\mathrm{L_X}$ > $10^{44}$ erg $\mathrm{s^{-1}}$. The number counts predicted from the integration of this XLF are also shown in the figure, with 1$\sigma$ uncertainties obtained from 10000 Monte Carlo realizations that propagate the parameter uncertainties reported for the PDE fit in Table 2 of \cite{pouliasis_active_2024}\footnote{This does not take into account the correlations of the parameters, that are presented in Fig. 7 of \citet{pouliasis_active_2024}. If these correlations were included, the result would be a broadening of the uncertainty band.}. These number counts show much better agreement with the Mock Catalogue, suggesting that the discrepancy between the mock catalogue and \cite{vito_high-redshift_2018} could be attributed to the limited number of bright AGN within the small survey area of CDFS. However, while this may explain the deviation at higher fluxes, it does not account for the discrepancy at the faintest fluxes.

Over the past two years, JWST has identified a significantly larger number of AGN candidates at high redshift, compared to previous observations.
The recent work by \citet{lyu2024active} presents a high redshift sample with 20 JWST selected AGN (candidates) and 2 pre-JWST selected AGN, at z > 4. If confirmed, this implies a tenfold increase in AGN detections, and with a survey area of 34 $\mathrm{arcmin^2}$, this would correspond to 2330 AGN\footnote{This estimate is also subject to considerable cosmic variance.} $\deg^{-2}$.

The higher number of AGN (candidates) identified by JWST suggests that current X-ray surveys could be missing a substantial fraction of the high redshift AGN population. The disparity between the predictions from the mock catalogue and the X-ray observations, at the faintest fluxes, may result from the limitations of current X-ray telescopes, which struggle to detect AGN at high redshifts. 

Recent JWST observations also find a higher black hole accretion rate density (BHARD; i.e. the average accretion rate, per comoving volume element, which quantifies the average growth of the AGN population), by a factor of $\sim$3, more consistent with the predictions from cosmological simulations \citep{yang2023ceers,pouliasis_active_2024}. Considering this, even though the predictions from the mock catalogue might be too optimistic, they offer valuable insight into an unexplored AGN population that recent JWST observations are beginning to unveil.

This comparison emphasizes the value of Cosmological Simulations of Galaxy Formation to assess the true abundance and properties of AGN, as well as to make predictions for future observatories (see also \citealt{habouzit_is_2024} for a recent comparison of cosmological simulations with results from JWST). Considering all of this, the mock catalogue appears to be robust to simulate the NewAthena survey.

\section{WFI survey simulation} \label{WFI Survey Simulation}

\subsection{SIXTE} \label{The SIXTE Simulation Tool}

In this work, we adopt the SIXTE software, a mission-independent Monte Carlo simulation package for X-ray observatories \citep{dauser_sixte_2019}. SIXTE was installed on Sciserver\footnote{https://www.sciserver.org/}, a collaborative research platform that provides access to large scientific datasets, computational tools, and cloud-based resources, allowing researchers to analyze and share data efficiently.

SIXTE operates in a fully modular approach, comprising three fundamental steps: photon generation, photon imaging, and event detection. The input source catalogue follows the conventions of the SIMPUT format \citep{schmid2013simput}, and is used to generate an initial photon list. This list contains the arrival time, energy, and position of each photon, which are determined through a Monte Carlo process based on the instrument's effective area, FOV, and attitude. This photon list is then convolved with the PSF and vignetting functions to generate an impact list of photons that reach the detector plane. Finally, this impact list is sent to the detector model, which simulates the event generation and readout process. The outcome of this process is a standardized event file that contains the read-out times, reconstructed energies, and detected positions of the photons. For further details, the reader is referred to \citet{dauser_sixte_2019}.

\subsection{Simulation process} \label{Simulation Process}

We assume that the WFI will be used in a wide survey comprising 100 pointings ($\sim44$ $\mathrm{deg^2}$). Of these, 70 pointings will have an exposure time of 200 ks, while the remaining 30 will have a deeper exposure of 300 ks.

Due to the time-consuming nature of these simulations, only 9 pointings are simulated: 6 with 200 ks of exposure and 3 with 300 ks of exposure. The results are then scaled to represent the complete survey by multiplying the number of detections by $\frac{70}{6}$ for 200 ks and $\frac{30}{3}$ for 300 ks observations. These pointings are non-contiguous and chosen to ensure maximum separation across different areas of the light cone, thus mitigating the impact of cosmic variance on our results. \citet{oogi2023uchuu} use a large volume simulation, to present an in depth discussion of the effect of cosmic variance in AGN surveys, finding that for an area of 10 $\mathrm{deg^2}$, the cosmic variance is < 10\%, even at z$\sim$8, for most of the AGN population, thus it should not significantly affect our results, except for the brightest sources (but see Section \ref{Discussion} for more on this).

The input catalogues include three types of source: AGN, non active galaxies, and extended sources. Since this work focuses on AGN, and other sources are treated as background contamination, using cosmological simulations to generate additional catalogues for them was deemed unnecessary. Therefore, the mock catalogues developed by \citet{marchesi_mock_2020} were utilized for the non active galaxies and extended sources. Although these mock catalogues account for nearly all of the CXB, we must still include the diffuse galactic foreground from the local hot bubble and diffuse galactic emission. This foreground was added as an additional SIXTE file, adopting the model presented in Appendix C of \citet{zhang_high-redshift_2020}. The instrumental background is assumed to be flat over the FoV and is directly implemented within SIXTE, using the most up-to-date files for NewAthena. The WFI instrument comprises four physically separated detectors, creating a cross-shaped pattern of insensitive regions in the resulting image. To mitigate these gaps and achieve a more uniform coverage, we assumed a dithering pattern following a Lissajous curve \citep{zhang_high-redshift_2020} with an amplitude of 0.025 degrees.

\begin{figure*}
\centering
\includegraphics[width=0.75\textwidth]{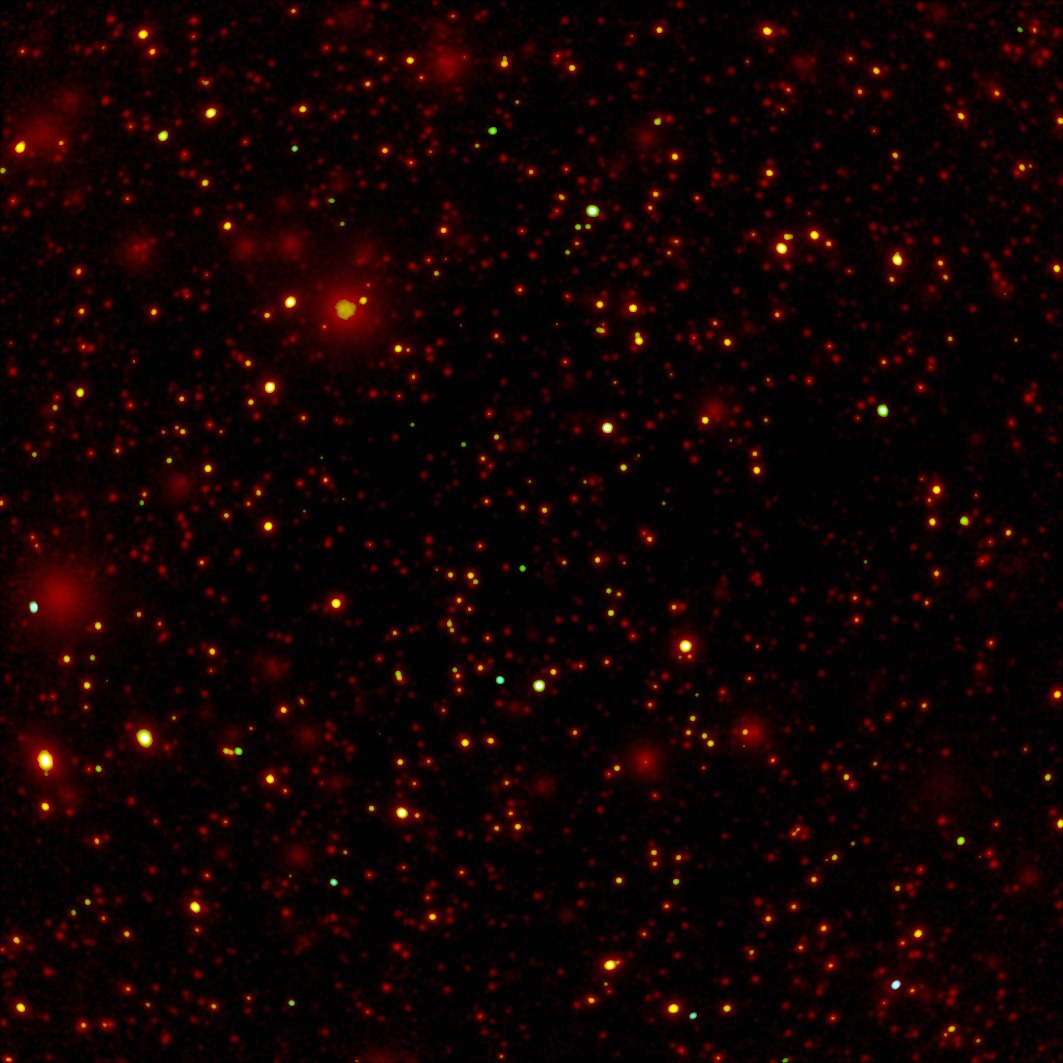}
\caption[RGB color image of a WFI FOV]{RGB color image of a WFI FOV (40'$\times$40'), with 200 ks of observation time. The red band corresponds to 0.5-2 keV, the green band to 2-4.5 keV, and the blue band to 4.5-10 keV. The image is on a logarithmic scale and is smoothed.}
\label{fig:Color_image}
\end{figure*}

Figure \ref{fig:Color_image} shows an example of an RGB image, of a WFI pointing, illustrating a future observation with NewAthena, with 200 ks of observation. The image shows several point sources, corresponding to AGN or non active galaxies, as well as extended sources, such as galaxy groups or clusters. AGN present a variety of colors, due to different spectral properties.

\subsection{Detection methodology} \label{Detection Methodology}

Source detection is performed in the soft band (0.5-2 keV), since this is the most sensitive band and ensures the best S/N for NewAthena, due to the lower effective area at higher energies. Furthermore, because of the varying exposures across different regions of the image (gaps and borders), each image is normalized by the corresponding exposure map before running the detection pipeline.

Source detection was performed using the Source Extractor software (SExtractor; \citealt{bertin_sextractor_1996}). SExtractor is a versatile software package originally developed for source detection in optical images. However, it has been successfully used in X-ray images, provided that they are appropriately filtered (e.g., \citealt{valtchanov_comparison_2001,pacaud_xmm_2006,zhang_high-redshift_2020}). The initial filtering step is crucial to denoise the image, thus reducing the number of spurious detections found by SExtractor. There are many approaches to denoise an astronomical image (see \citealt{roscani_comparative_2020} for a recent work comparing different denoising algorithms) and wavelet filtering is a very effective approach for X-rays.

The denoising process involves three steps: first, the image is subjected to a wavelet transform that decomposes it into various frequency components represented by wavelet coefficients. Second, thresholding is applied, reducing or setting to zero the smaller coefficients likely representing noise and retaining the larger coefficients that correspond to signal features. Lastly, an inverse wavelet transform is applied to reconstruct the data, yielding a denoised image, where noise is suppressed and essential features are preserved. This algorithm was implemented using the Python routine: "$\mathrm{denoise\_wavelet}$" from the scikit-image\footnote{https://scikit-image.org/} library.

\begin{figure}
\centering
\includegraphics[width=\columnwidth]{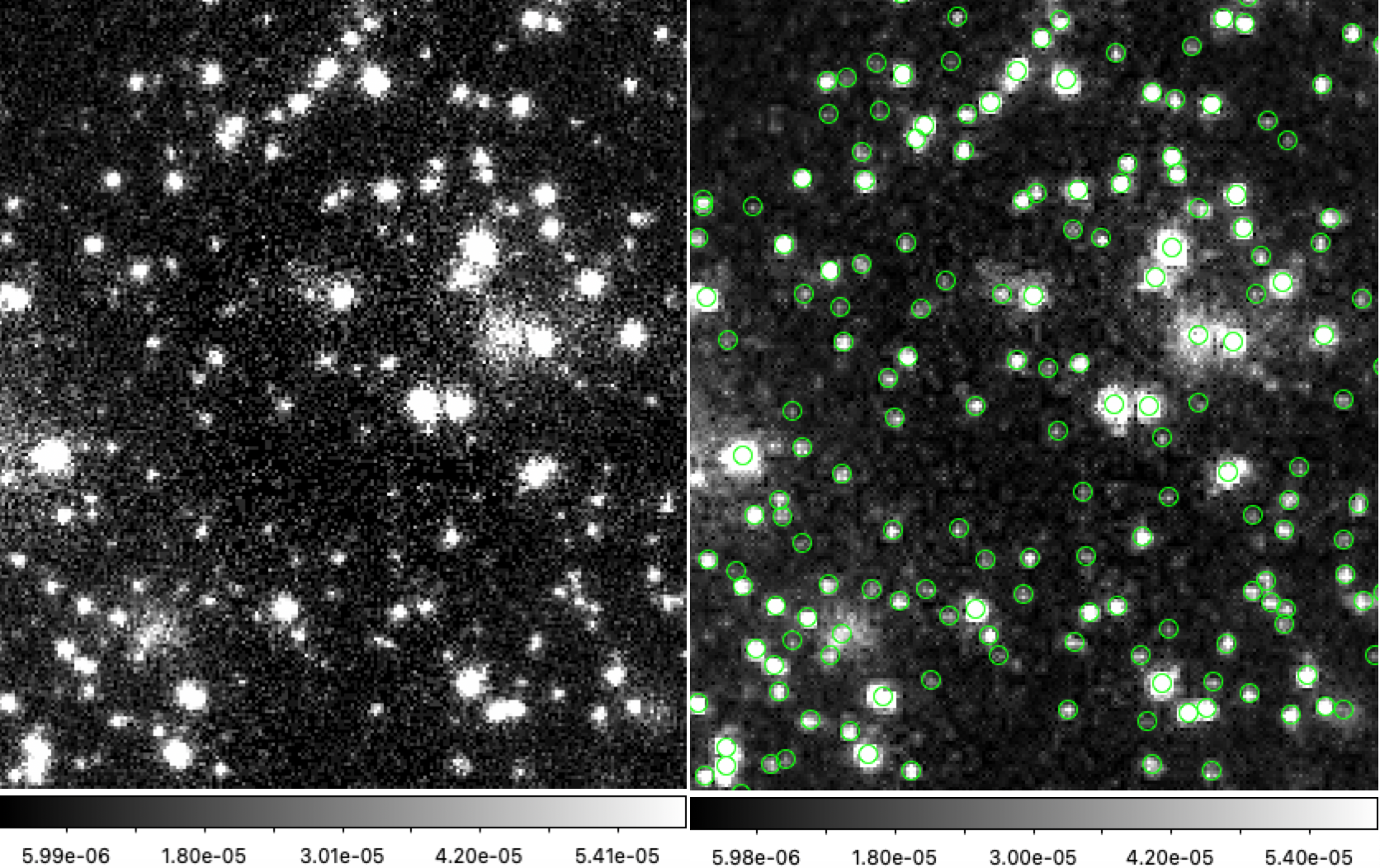}
\caption[Zoom of a simulated WFI FOV]{Comparison of unfiltered (left panel) and filtered (right panel) views of a simulated Athena/WFI zoomed field of view (9'$\times$9', 300 ks exposure, 0.5-2 keV soft band). The count rate is plotted in the color bar. The green circles are the sources detected by SExtractor.}
\label{fig:Filtered_image}    
\end{figure}

To calibrate the denoising algorithm, the framework developed by \citet{batson_noise2self_2019} was adopted. This approach focuses on minimizing the self-supervised loss\footnote{It is called self-supervised since it does not require knowledge of the ground truth.}, which quantifies the average difference between filtered and noisy images. Under the assumption that the noise exhibits statistical independence, while the true signal exhibits some correlation, this metric enables an objective comparison between different filters, based solely on the noisy data. Using this criterion, all discrete wavelets\footnote{Other algorithms available in the library, namely: the bilateral filter, non-local means, and total variance were also tested, however, none of these was successful.} from the PyWavelets\footnote{https://pywavelets.readthedocs.io/en/latest/} library were tested. The best results were obtained with the wavelet $coif1$ of the coiflet family and 1 decomposition level\footnote{Using more than 1 decomposition level reduces the positional accuracy of the detected sources.}. After choosing this wavelet, the calibration process was repeated to select the best noise standard deviation, used to calculate the thresholds, for each image. Fig. \ref{fig:Filtered_image}, compares the unfiltered and filtered versions of a (zoomed) image, with 300 ks of observation. As shown in the figure, the noise is suppressed without significant blurring of the sources. 

Following this preprocessing step, SExtractor was applied to each image.
The objects were selected with a 3$\sigma$ threshold, per pixel, on a minimum area of 3 contiguous pixels. These objects were deblended assuming 64 levels, with a minimum contrast of 0.005. The background map was also used as a weight map, to account for the different exposures, across different regions of the image. Photometry was performed using SExtractor's automatic aperture photometry, following \citet{kron_photometry_1980} and \citet{infante_faint_1987}. Although source detection was conducted on the filtered image, the photometry was extracted from the original image. 

After extracting the sources, they are crossmatched with the input catalogues. This crossmatch was performed using TOPCAT\footnote{https://www.star.bris.ac.uk/~mbt/topcat/}, assuming a maximum separation of 5" and selecting as valid cross-matches those that were symmetric. We matched the output catalogue with the 3 input catalogues and in cases where the crossmatch radius was compatible with both an AGN and a galaxy, the source with the highest flux was taken as the counterpart. Sources without match in the input catalogues are classified as spurious. A flux limit\footnote{This is the predicted (soft band) confusion limit for a 9" PSF.} of $5\times10^{-17}$ $\mathrm{erg}$ $\mathrm{s^{-1}cm^{-2}}$ was applied to the input catalogues, to mitigate the possibility of coincidence matching of faint sources with spurious background fluctuations. 

\subsubsection{Completeness and reliability} \label{Completeness and Reliability}

With this pipeline, we measure, on average, a 4\% spurious source fraction in the survey. Specifically, this fraction is expected to be 4.4\% in pointings with 200 ks of exposure and 3.1\% with 300 ks of exposure. The image borders have a much lower exposure, contributing to a larger spurious fraction. If these are excluded, retaining 90\% of the central area, the average spurious fraction reduces to 3.0\% (3.2\% for 200 ks and 2.6\% for 300 ks, respectively).

\begin{figure}
  \centering
  \begin{subfigure}{0.495\linewidth}
    \centering
    \includegraphics[width=\linewidth, height = 45mm]{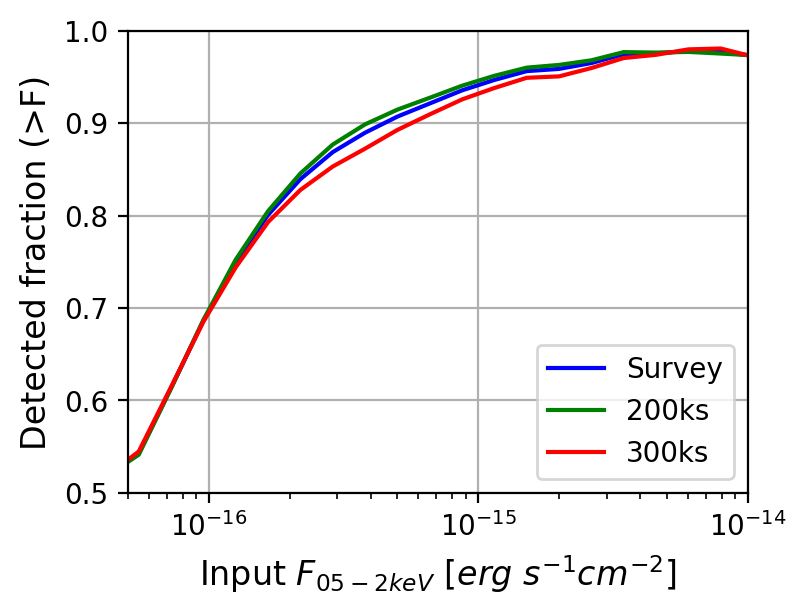}
  \end{subfigure}
  \hfill
  \begin{subfigure}{0.495\linewidth}
    \centering
    \includegraphics[width=\linewidth, height = 45mm]{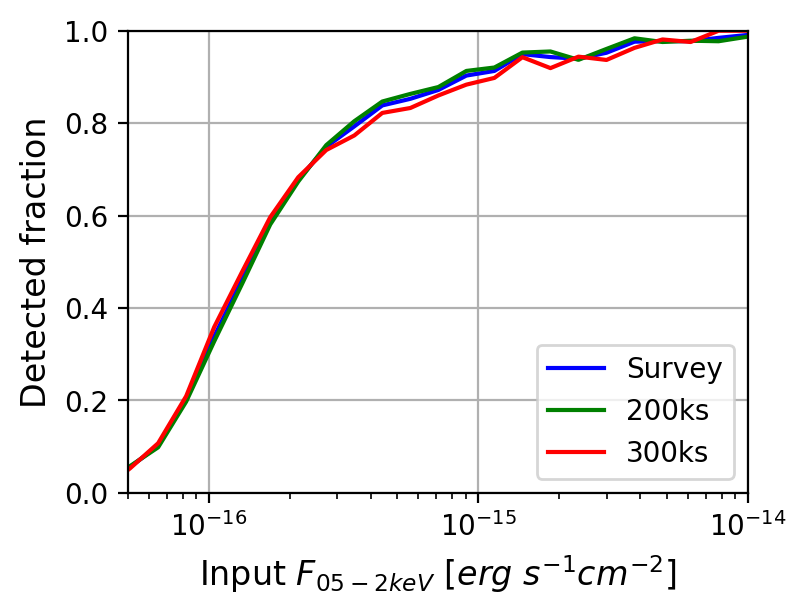}
  \end{subfigure}
  \caption[Completeness curves]{Left: Cumulative completeness curves as a function of input flux ($F_{0.5-2 \text{keV}}$) for different survey exposure times. The blue curve represents the baseline survey, while green and red correspond to 200ks and 300ks exposures. Right: The same as the left panel, but binned in flux.}
  \label{fig:completeness}
\end{figure}

Next, we consider the completeness of the survey, defined as the fraction of input sources successfully detected. This is plotted as a function of the input soft band flux, shown cumulatively in the left panel of Fig. \ref{fig:completeness} and differentially, in the right panel of the same figure. The fraction of detected sources is greater than 95\% for high fluxes, above $10^{-15}$ erg $\mathrm{s^{-1} cm^{-2}}$, dropping to around 50\%, for fluxes above $5\times10^{-17}$ erg $\mathrm{s^{-1} cm^{-2}}$.  The 80\% completeness flux limit for this survey is $3.8\times10^{-16}$ erg $\mathrm{s^{-1} cm^{-2}}$, while the 20\% completeness flux limit is $8.8\times10^{-17}$ erg $\mathrm{s^{-1} cm^{-2}}$. The figure also shows the completeness in the pointings with 200 ks (green line) and 300 ks (red line) of exposure. Visual inspection of undetected sources indicates that the primary cause of incompleteness, at high flux levels (i.e., above $10^{-15}$ erg $\mathrm{s^{-1} cm^{-2}}$), is source blending, predominantly with other AGN. Nonetheless, extended sources also contribute significantly to the incompleteness at fainter fluxes. This is evident in the 300 ks completeness curve, which falls below the 200 ks curve within the flux range of $2 \times 10^{-16}$ to $2 \times 10^{-15}$ erg $\mathrm{s^{-1}~cm^{-2}}$. The reduction in completeness in the deeper exposure is primarily due to the presence of two bright clusters within one of the simulated pointings.

\begin{figure}
\centering
\includegraphics[width=0.8\columnwidth]{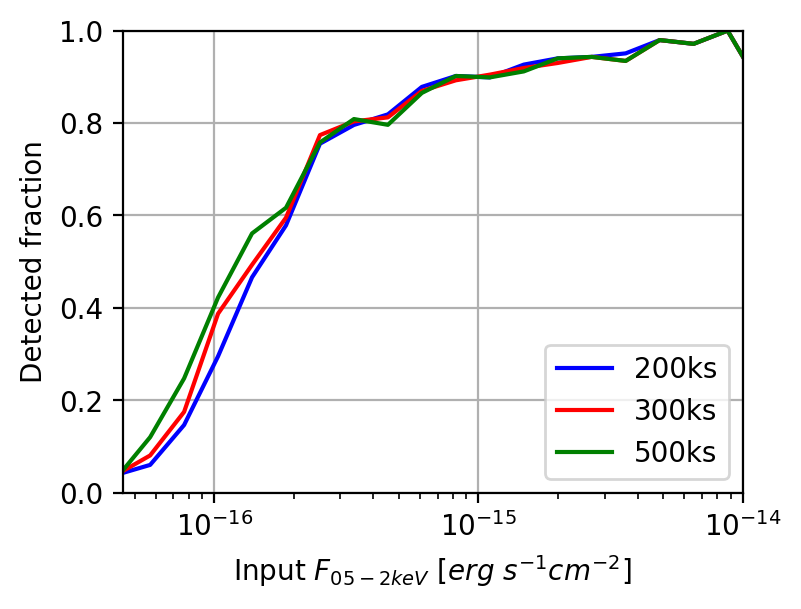}
\caption[Completeness Comparison plot]{The plot compares the completeness as a function of input soft band flux (0.5–2 keV) for three different exposure times: 200 ks (blue line), 300 ks (red line) and 500 ks (green line), in the same FOV. Deeper exposures consistently show higher detection fractions, at lower flux values, reflecting the improved sensitivity of longer exposure times.}
\label{fig:Compleness_comparison}    
\end{figure}

Surprisingly, the completeness seems to be independent of the exposure of the survey. This is unexpected since increasing the exposure time increases the S/N, thus it should improve the completeness for faint sources. To further investigate this, Fig. \ref{fig:Compleness_comparison} plots the completeness curves for the same FOV, with 200 ks of exposure (blue line), 300 ks of exposure (red line) and 500 ks of exposure (green line). The figure demonstrates that deeper observations achieve better completeness for faint fluxes. The absence of this trend in the previous figure is probably due to the cosmic variance among different pointings of the WFI. The completeness curves converge at $5\times10^{-17}$ erg $\mathrm{s^{-1} cm^{-2}}$, which indicates that below this flux, the survey is confused and deeper exposures will not improve the depth.

\subsubsection{Astrometric and Photometric accuracy} \label{Astrometric and Accuracy}

\begin{figure}
\centering
\begin{subfigure}{0.495\linewidth}
    \centering
    \includegraphics[width=\linewidth, height = 40mm]{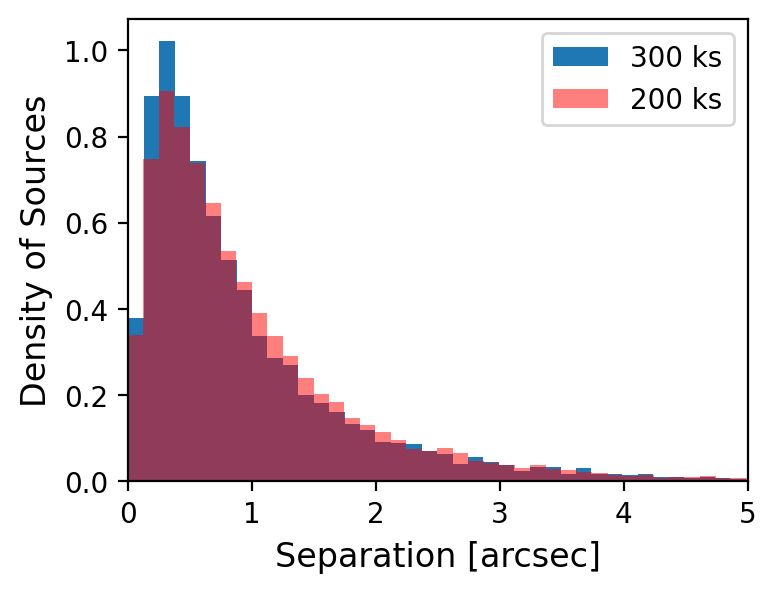}
    \label{fig:Separation}
\end{subfigure}
\hfill
\begin{subfigure}{0.495\linewidth}
    \centering
    \includegraphics[width=\linewidth, height = 40mm]{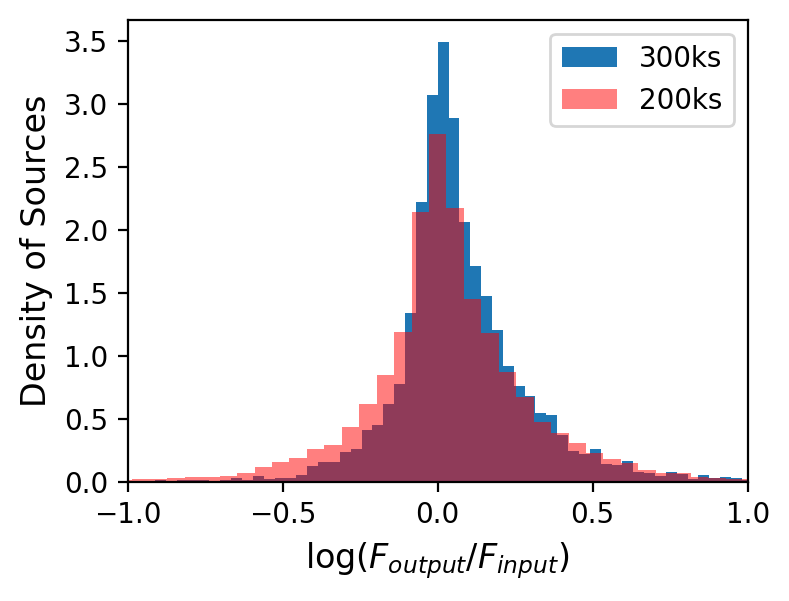}
    \label{fig:flux_ratio}
\end{subfigure}
\caption[Source detection accuracy] {Left Panel: Histogram of AGN separations between input and detected positions, in arcseconds, comparing data with 200 ks (red) and 300 ks (blue) exposures. The distributions show the normalized frequency of separations between detected sources. Right Panel: Histogram of the normalized source fraction versus the logarithmic flux ratio, $\log(F_{\text{output}}/F_{\text{input}})$, for 200 ks (blue) and 300 ks (red) exposures. The distribution is centered around zero, indicating accurate flux recovery, with the 300 ks exposure reducing the negative tail, suggesting a more accurate flux recovery for longer exposures.}
\label{fig:detection_accuracy}
\end{figure}

Another important factor to evaluate is the positional accuracy of the AGN, which is crucial to ensure that multiwavelength counterparts can be found. The left panel of Fig. \ref{fig:detection_accuracy} plots the difference between the input and detected positions of AGN observed with 200 ks (red histogram) or 300 ks (blue histogram). The median separation of detected AGN is 0.64" for the 300 ks exposure and 0.71" for the 200 ks exposure. Additionally, the fraction of AGN with separations less than 1" is 68\% for 300 ks and 65\% for 200 ks. 

Finally, we assess the accuracy of the detected fluxes. To calibrate the measurements, we use bright AGN with fluxes above $10^{-14}$ erg $\mathrm{s^{-1} cm^{-2}}$, deriving a mean energy conversion factor of $2.5\times10^{-13}$ erg $\mathrm{cm^{-2} cts^{-1}}$. The right panel of Fig. \ref{fig:detection_accuracy} presents the ratio of detected to input flux for sources with $\mathrm{log(N_H)} \leq 22$, comparing observations with 200 ks (blue histogram) and 300 ks (red histogram). While the flux ratios are centered around unity, they exhibit significant scatter. The 300 ks exposure results in a narrower negative tail, suggesting that deeper observations mitigate flux underestimation. Quantitatively, 69\% of AGN in 300 ks exposures have detected fluxes within 50\% of their true values, compared to 62\% for 200 ks. Additionally, 77\% of AGN exhibit flux ratios below 1.5, independently of exposure time. These findings suggest that flux underestimation primarily arises from the low photon counts of faint AGN, whereas overestimation is likely driven by contamination from nearby sources.

\section{Results} \label{Results}

Each pointing of the WFI will detect $\sim$2800 point sources, on average, with 90.7\% identified as AGN and the remaining 9.3\% as non-active galaxies.
The complete survey will find a total of $\sim$250\,000 AGN, including $\sim$9000 CTK AGN. The predictions for this survey are presented in Tab. \ref{tab:survey}, while the detected sources are plotted in Fig. \ref{fig:high_z}.

\begin{table*}
\caption[Predictions for the Survey of the NewAthena mission.]{Predictions for the Survey of the NewAthena mission. These results are extrapolated from 9 pointings taken from a grid in the mock catalogue. The different columns show a breakdown of the number of detected AGN above a given redshift threshold, as well as the logarithm of the 0.5-2 keV Luminosity, assuming a flux of $5\times10^{-17}$ $\mathrm{erg s^{-1} cm^{-2}}$. The first row presents the results for all sources, while the second row only considers Compton Thick AGN.}
\centering
\resizebox{\textwidth}{!}{  
\begin{tabular}{c cc cc cc cc cc cc cc}
\hline
 & \multicolumn{2}{c}{z\textgreater{}1} & \multicolumn{2}{c}{z\textgreater{}2} & \multicolumn{2}{c}{z\textgreater{}3} & \multicolumn{2}{c}{z\textgreater{}4} & \multicolumn{2}{c}{z\textgreater{}5} & \multicolumn{2}{c}{z\textgreater{}6} & \multicolumn{2}{c}{z\textgreater{}7} \\ 
 & \multicolumn{1}{c}{$N_{src}$} & $L_{lim}$ & \multicolumn{1}{c}{$N_{src}$} & $L_{lim}$ & \multicolumn{1}{c}{$N_{src}$} & $L_{lim}$ & \multicolumn{1}{c}{$N_{src}$} & $L_{lim}$ & \multicolumn{1}{c}{$N_{src}$} & $L_{lim}$ & \multicolumn{1}{c}{$N_{src}$} & $L_{lim}$ & \multicolumn{1}{c}{$N_{src}$} & $L_{lim}$ \\ 
\hline
All AGN & 187505 & 41.4 & 86237 & 42.2 & 21046 & 42.6 & 3190 & 42.8 & 435 & 43.0 & 35 & 43.2 & 11 & 43.4 \\ 
CTK AGN & 7442 & 43.3 & 5335 & 43.7 & 1601 & 44.0 & 193 & 44.2 & 11 & 44.3 & 0 & - & 0 & - \\ 
\hline
\end{tabular}
}
\label{tab:survey} 
\end{table*}

As shown in Tab. \ref{tab:survey}, NewAthena is expected to detect $\sim$21\,000 AGN at z>3 and $\sim$3\,200 AGN at z>4. This is unprecedented among X-ray surveys, as the largest high redshift sample, compiled by \citet{pouliasis_active_2024}, includes $\sim$630 z>3 and $\sim$100 z>4 AGN, implying an improvement by a factor of $\sim$30. NewAthena will also enable the detection of a population of X-ray-selected AGN within the Epoch of Reionization (EoR), at z>6. Currently, most of z>6 AGN observed in the X-ray were pre-selected in the Optical or Near Infrared (NIR) \citep{wang_revealing_2021} and blind X-ray detection is necessary to statistically constrain the properties of the AGN population \citep{pouliasis_active_2024}\footnote{If pre-selection is used, the survey's selection function becomes unclear, leading to potential biases in the conclusions drawn.}. Recently, a few X-ray selected AGN were found at z>6, \citep{wolf_first_2021,wolf_x-ray_2023,barlow-hall_constraints_2023}, however, this is only possible for the most luminous sources ($L_{2-10keV}$>$10^{45}$ erg $s^{-1}$), and current facilities are unable to detect the bulk of the AGN population. NewAthena is expected to uncover X-ray sources 2 orders of magnitude fainter.

The source density derived by \citet{marchesi_mock_2020} sets a good comparison point for our estimates since they were derived for the original ATHENA configuration and are purely based on observational constraints (XLF+CXB). \citet{marchesi_mock_2020} predict 8220 AGN with z>3, 1695 for z>4, 385 for z>5, 75 AGN for z>6 and 21 for z>7\footnote{These numbers are rescaled to 100 WFI pointings, since \citet{marchesi_mock_2020} assumed 108 WFI pointings in their Mock Observing Plan.}. Our predictions diverge significantly for AGN at redshifts z>3 and z>4, where our results are higher by factors of 2.6 and 1.8, respectively. Although there is a reasonable agreement for AGN at z>5, our predictions are lower by factors of 2.4 at z>6 and 1.9 at z>7, in the EoR. Furthermore, considering the predictions for the CTK population, \citet{marchesi_mock_2020} estimates 6700 AGN, with 219 at z>3. In contrast, our predictions for CTK AGN at z>3 are higher by a factor of 7. This is further discussed in Sect. \ref{Discussion}.

\begin{figure}
\centering
\includegraphics[width=0.8\columnwidth]{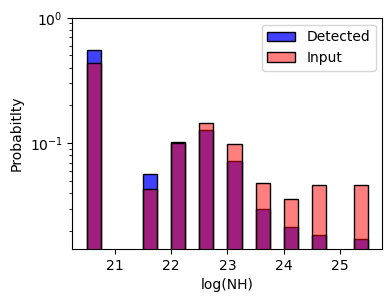}
\caption[Column density]{Distribution of column density for input (red) and detected (blue) AGN. The input AGN have a flux limit of $5\times10^{-17}$ $\mathrm{erg}$ $\mathrm{s^{-1} cm^{-2}}$.}
\label{fig:NH}    
\end{figure}

\begin{figure}
\centering
\includegraphics[width=\columnwidth]{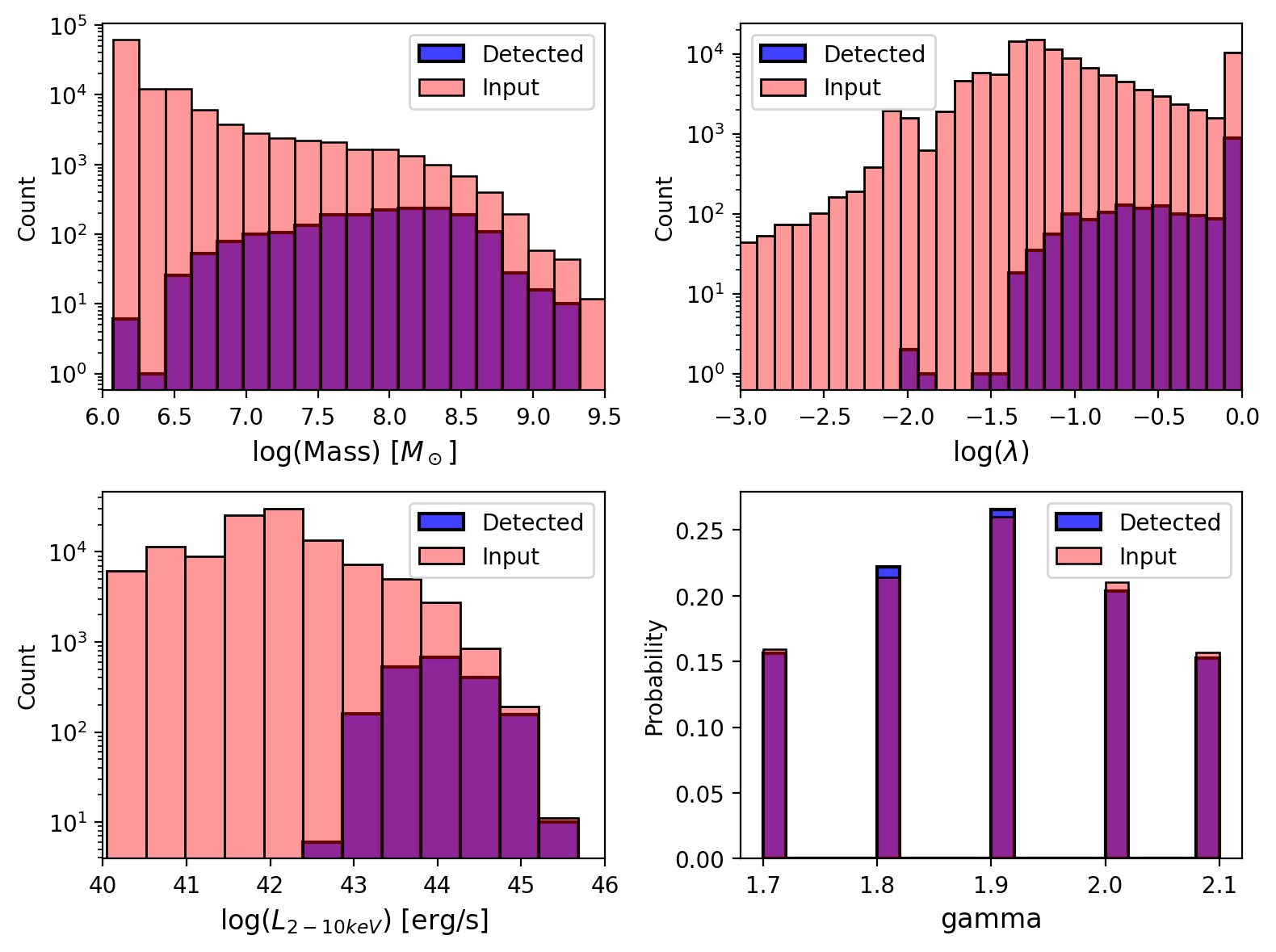}
\caption[Comparison of input and detected AGN at z > 3]{Comparison of input and detected AGN at z > 3. The histograms show the distributions of SMBH mass (upper left panel), Eddington ratio ($\lambda$) (upper right panel), X-ray luminosity ($L_X$) (lower left panel), and photon index ($\Gamma$) (lower right panel) for both input (red) and detected (blue) populations. Detected sources tend to have higher masses, X-ray luminosities, and Eddington ratios, with similar distributions for $\Gamma$.}
\label{fig:high_z}    
\end{figure}

Fig. \ref{fig:NH} compares the $N_H$ distribution of detected and input AGN. The figure demonstrates that NewAthena will be capable of detecting highly obscured AGN, including those with $N_H$ > $10^{25}$ $cm^{-2}$, however, a strong selection bias remains against the most obscured AGN.

One advantage of creating catalogues with HDS is that these simulations track some intrinsic characteristics of the AGN, such as the SMBH mass or accretion rate. Fig. \ref{fig:high_z} illustrates the normalized distributions of SMBH masses, Eddington ratios ($\lambda$), hard X-ray luminosities, and photon indices ($\Gamma$) of the (detected) high redshift AGN (z>3), as well as the input AGN\footnote{No flux limit is assumed in the input distribution.}. The figure shows that, according to IllustrisTNG, most of the high redshift AGN will be powerful quasars, with masses $\sim$$10^8$ $\mathrm{M_\odot}$, accreting very close to the Eddington limit, with a typical luminosity of $\sim$$10^{44}$ erg $s^{-1}$. In addition, NewAthena might be able to uncover some AGN with SMBH masses as low as $\sim$$5\times10^6$ $\mathrm{M_\odot}$. When comparing the input and detected populations, it becomes evident that low-mass SMBHs power the majority of high redshift AGN and highlights that the observed sample is not representative of the input population. The distribution of Eddington ratios spans a wide range, with the majority of SMBHs accreting at a small fraction of the Eddington limit, typically a few percent, consistent with an ADAF scenario. Nevertheless, AGN accreting near the Eddington limit constitute a considerable fraction of the input population, indicating a diverse range of accretion states. The photon indices present a similar distribution, indicating that there is no bias in this quantity.

\section{Discussion} \label{Discussion}

As stated in \citet{marchesi_mock_2020}, one of the goals for the Mock Observing Plan of the original ATHENA was to detect at least 10 AGN at 7<z<8 with 44<$\mathrm{log(L_{0.5-2keV})}$<44.5. According to our predictions, NewAthena will also achieve this goal. No AGN at z > 6 with $\mathrm{log(L_{0.5-2keV})} < 43.5$ was detected in the simulated survey area (9\% of the total), despite the input catalogue containing 8 of these AGN. This suggests that significantly deeper exposures would be necessary to detect such sources. 

Since NewAthena is a rescoped version of ATHENA, it is insightful to compare our predictions with the predictions for the original wide survey. As discussed in the previous section, the predictions for AGN in the redshift range 3<z<5 are higher by a factor of $\sim$2. This increase can be attributed to the higher number of high redshift AGN in the mock catalogue, as illustrated by Fig. \ref{fig:number_counts_high_z}, compared to the mocks used by \cite{marchesi_mock_2020}. 

Notably, the predictions for the EoR are approximately half of what is predicted by \citet{marchesi_mock_2020}. This discrepancy is probably due to the higher levels of obscuration assumed in our mock. While \citet{marchesi_mock_2020} adopt a constant CTK fraction of $\frac{4}{9}$, the mock catalogue employs a redshift-and-luminosity-dependent CTK fraction. In addition, ISM obscuration is taken into account, assuming the model of \citet{gilli_supermassive_2022}, in which the median column density of the host galaxy becomes CTK at z>6. 

Finally, it is necessary to acknowledge that HDS face substantial difficulties in reproducing the most luminous AGN ($L_X$>$10^{45}$ erg $\mathrm{s^{-1}}$) observed at high redshift, possibly due to limitations in volume and resolution \citep{weinberger_supermassive_2018,amarantidis_first_2019}. Specifically, IllustrisTNG is unable to simulate SMBHs more massive than $5\times10^7$ $\mathrm{M_{\odot}}$ at z>6, and assumes Eddington limited accretion, which restricts the maximum attainable luminosity. For this reason, the number of detected EoR AGN could be higher than these predictions.

Nevertheless, these findings are encouraging, as they demonstrate NewAthena's capability to detect a population of X-ray selected AGN during the EoR, even with the revised mission specifications and higher levels of obscuration.

\subsection{CTK AGN detection} \label{CTK AGN detection}

Regarding the detection of CTK AGN, the total number of these sources significantly exceeds the predictions by \citet{marchesi_mock_2020}, with a discrepancy by a factor of 7 at z>3\footnote{By definition, only AGN with $\mathrm{N_H} \geq 1.5\times10^{24}$ $\mathrm{cm^{-2}}$ are classified as CTK, so only AGN with $\mathrm{log(N_H)} \geq 24.18$ are considered CTK.}. This increase can be attributed to a higher overall number of AGN in the mock catalogue and the assumption of a CTK fraction that increases with redshift. This aligns well with the most recent results from JWST, which suggest a greater prevalence of CTK AGN at high redshifts \citep{maiolino_jwst_2024,lyu2024active}, however, large uncertainties remain since JWST is unable to directly constrain the column densities and X-ray luminosities of these X-ray weak AGN.

If our predictions hold, NewAthena will detect a significant number of CTK AGN until z$\sim$4 and may even detect a few sources at z>5. This would represent a key advancement in understanding these rare, highly obscured sources at high redshifts. Detecting CTK AGN at such early times could provide essential clues about the formation and growth of SMBH, as well as the role of these objects in the evolution of galaxies. This is discussed further in Sec. \ref{Detection Methodology}.

\subsection{Comparison with theoretical estimates} \label{Comparison with theoretical estimates}

Our predictions are 2 dex lower than those presented in \citet{habouzit_supermassive_2021} with IllustrisTNG, across all redshifts. This discrepancy is attributed to differences in the modeling of X-ray properties, including the bolometric luminosity model, radiative efficiency, and bolometric corrections, as well as the higher obscuration assumed in this study. Additionally, realistic Monte Carlo simulations were conducted, accounting for background and foreground contaminants, and the AGN were detected using a source detection pipeline, instead of applying a simple luminosity limit.

This significant discrepancy highlights the critical role that post-processing assumptions play in the creation of mock catalogues and predictions for future missions. As discussed in Sec. \ref{Mock catalogue validation}, accurate modeling of the X-ray properties, down to the faintest AGN, is essential to align the predictions of cosmological simulations with observational constraints, such as the CXB, and ensure robust estimates.

\subsection{Survey observing strategy and limitations} \label{Detection Methodology}

Comparing the survey performance between pointings with 200 ks and 300 ks of exposure reveals minimal gains in completeness, as shown in Fig. \ref{fig:completeness}, with the 300 ks pointings detecting only an average of 100 additional AGN. However, a deeper exposure increases the number of counts per source, resulting in more accurate estimates of properties such as flux, as demonstrated in Fig. \ref{fig:detection_accuracy}. Importantly, all detected AGN at z > 6 were found in 200 ks exposures, suggesting that this is sufficient to reach the EoR, until z$\sim$8. Given the significant time investment required for the deeper exposures (3 Ms), a rigorous evaluation of whether the increase in S/N improves the overall scientific return of the survey is essential. 

The main limitation of this work is the reliance on a single cosmological simulation: IllustrisTNG. As discussed in \citet{amarantidis_first_2019} and \citet{habouzit_supermassive_2021}, different simulations employ distinct subgrid modeling, which substantially affects the evolution and properties of SMBH, thus producing different AGN populations. Consequently, the results presented here may not be fully representative of the theoretical understanding of AGN.

This limitation can be addressed in future work, comparing the results of mock catalogues from different cosmological simulations, with several promising candidates among the state-of-the-art HDS.  For instance, the MillenniumTNG simulation
\citep{pakmor_millenniumtng_2023,kannan_millenniumtng_2023}; the FLAMINGO simulation \citep{schaye_flamingo_2023}; the Astrid simulation \citep{ni_astrid_2022, ni_astrid_2024} or the SIMBA simulation \citep{dave_simba_2019}.

An additional limitation concerns the source detection and extraction methodology. Although the current methodology ensures high completeness, reliability, and astrometric accuracy, it fails to accurately extract counts from sources, especially if they are blended or contaminated. 

An important improvement to the source detection pipeline would be to implement model-fitting capabilities, incorporating the PSF and extended source models. Such an improvement would not only increase the photometric accuracy by optimizing the S/N but would also enable more accurate measurements of blended or contaminated sources. Furthermore, this approach would enable the calculation of detection likelihoods, ultimately improving the reliability of source detections.

Future work could address this by adapting existing software, such as the $emldetect$ from the XMM-SAS package\footnote{https://www.cosmos.esa.int/web/xmm-newton/sas} or explore alternative approaches (e.g., machine learning; \citealt{xu_surveying_2024}). 

Finally, while our predictions suggest the detection of a significant number of CTK AGN, we have not evaluated whether their column density can be accurately constrained, which is crucial for their proper characterization. Future work should perform spectral analysis of the detected AGN to address this limitation and assess NewAthena’s ability to distinguish between intrinsically X-ray weak and CTK AGN. Additionally, it is important to investigate the impact of exposure time on X-ray spectra, as observations with 500 ks remain feasible despite the effects of source confusion.

\section{Summary and conclusions} \label{Summary and conclusions}

This work focused on simulating NewAthena´s capabilities to detect faint AGN. Using the IllustrisTNG cosmological simulation, a comprehensive mock catalogue of X-ray AGN was created and processed through an end-to-end simulation framework, with the SIXTE software. The analysis explored key aspects such as the completeness, reliability, and sensitivity of the survey, with a particular emphasis on the detection of high redshift and CTK AGN. The results demonstrated that NewAthena will significantly improve the detection of faint AGN populations, compared to current surveys, providing the first statistically significant X-ray selected sample of AGN in the EoR and advancing our understanding of AGN evolution in the early universe.

Our main findings can be summarized as:

\begin{itemize}
    \item A state-of-the-art hydrodynamical cosmological simulation, IllustrisTNG, was used to create the X-ray AGN mock catalogue that spans 10 $\mathrm{deg^2}$ and covers the redshift range 0-12. This involved the construction of a light cone from the discrete snapshots and applying corrections to account for the limited resolution of the simulation. These corrections have a significant impact on high redshift predictions. 

    \item We determined the X-ray properties of the SMBH in post-processing, assuming a bolometric luminosity model, X-ray bolometric corrections, an obscuration model and realistic spectral templates. We also demonstrate that the radiative efficiency and the low Eddington ratio AGN have a significant impact on the number counts, emphasizing the necessity to integrate subgrid prescriptions to track the SMBH spin in future large-scale cosmological simulations.

    \item During the validation process, we noticed an overabundance of faint AGN. However, this discrepancy can be alleviated by the assumption of a high CTK fraction and intrinsic X-ray weakness, suggested by recent JWST observations.

    \item We simulate the NewAthena survey and prove that it will significantly improve the statistics of high redshift AGN (z>3), and uncover a population of X-ray selected AGN in the EoR. Nevertheless, only AGN with luminosities exceeding $10^{43.5}$ erg $s^{-1}$ are expected at these redshifts. Our results also suggest a higher number of CTK AGN than previously predicted, even beyond z>4.

    \item The analysis of WFI pointings with varying exposure times indicates that deeper observations with 500 ks remain feasible, despite the challenges posed by source confusion. 

\end{itemize}

In summary, this work contributes to our understanding of high redshift AGN populations, emphasizing the implications for the NewAthena mission.

\section*{Acknowledgements}
We thank the referee for the helpful comments and ideas that improved this manuscript. NC, IM and JA acknowledge financial support from the Science and Technology Foundation (FCT, Portugal) through research grants UIDB/04434/2020 (DOI: 10.54499/UIDB/04434/2020) and UIDP/04434/2020 (DOI: 10.54499/UIDP/04434/2020). GL, SM, and AC acknowledge financial support from the European Union’s Horizon 2020 Programme under the AHEAD2020 project (grant agreement n. 871158) and the "Accordo Attuativo ASI-INAF" n. 2019-27-HH.0. PP was supported by FCT Principal Investigator contract
CIAAUP-092023-CTTI

\section*{Data Availability}
IllustrisTNG data is available at: https://www.tng-project.org/



\bibliographystyle{mnras}
\bibliography{Tese2} 




\begin{appendix}
     
\section{Resolution Correction for the Eddington ratio} \label{Resolution Correction for the Eddington ratio}

\begin{figure}
  \centering
  \begin{minipage}{\columnwidth} 
    \centering
    \begin{subfigure}{0.495\columnwidth} 
      \centering
      \includegraphics[width=\linewidth]{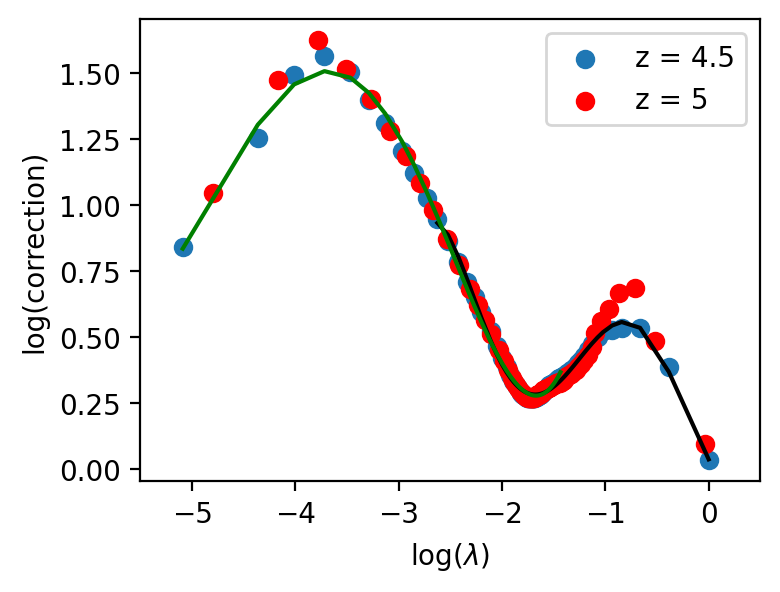}
    \end{subfigure}%
    \begin{subfigure}{0.495\columnwidth} 
      \centering
      \includegraphics[width=\linewidth]{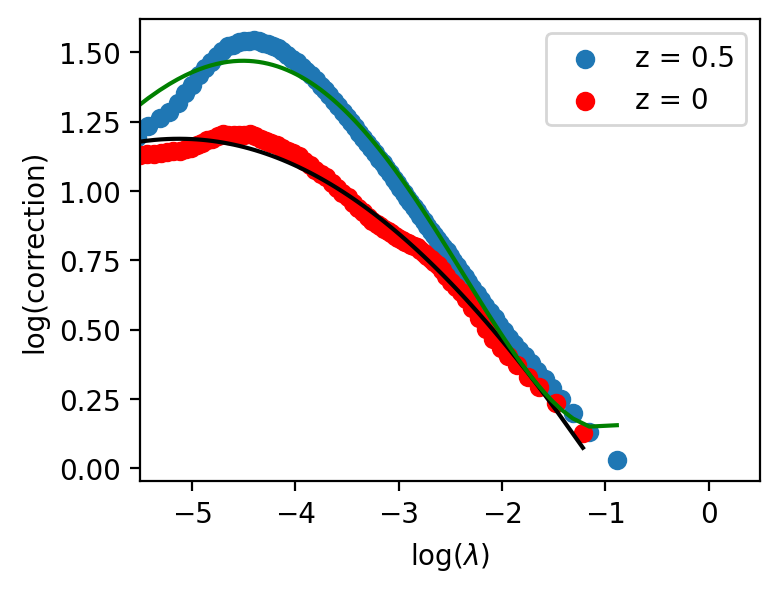}
    \end{subfigure}
    \caption{Left panel: The logarithm of the correction against the logarithm of the Eddington ratio, $\lambda$, for two redshifts: z = 4.5 (blue points) and z = 5 (red points). The smooth curves fit the data points at z = 4.5, green for log($\lambda$) <-2 and black for log($\lambda$) >=-2. The correction at z = 4.5 is very similar to the correction at z = 5. Right panel: The same as the left panel, but for z = 0 and z = 0.5.}
    \label{fig:correction_figure}
  \end{minipage}
\end{figure}

The resolution corrections were calculated as follows: for each of the 11 snapshots (z = 0 to z = 5, $\Delta z = 0.5$), all SMBH in TNG100-1 (1.6$\times10^6$ $\mathrm{M_\odot}$ baryonic mass resolution) and TNG100-2 (1.1$\times10^7$ $\mathrm{M_\odot}$) were retrieved. Then both distributions were binned by frequency in 100 bins. For each bin, the mean value was calculated and the ratio between the TNG100-1 and TNG100-2 values was plotted as a function of the values in TNG100-2. Subsequently, polynomial fits were applied to the data to capture the trends in the corrections (see Figure \ref{fig:correction_figure} for an example).

\begin{onecolumn}

\begin{longtable}{|c|c|c|c|c|c|c|}
    \captionsetup{width=0.9\textwidth} 
    \caption{Polynomial corrections for the Eddington ratios of the TNG300-1 simulation. The polynomial is given by $\log_{10}(c) = ax^5 + bx^4 + cx^3 + dx^2 + ex + f$, where $x = \log_{10}\left(\frac{\dot{M}}{\dot{M_{Edd}}}\right)$ is the logarithm of the Eddington ratio. For most redshift intervals, two separate polynomials are used: one for high Eddington ratios ($x \geq -2$) and another for low Eddington ratios ($x < -2$).} \label{tab:resolution} \\

    \hline
    \textbf{Redshift Range} & \textbf{a} & \textbf{b} & \textbf{c} & \textbf{d} & \textbf{e} & \textbf{f} \\ \hline
    \endfirsthead

    \caption[]{(continued)} \\

    \hline
    \textbf{Redshift Range} & \textbf{a} & \textbf{b} & \textbf{c} & \textbf{d} & \textbf{e} & \textbf{f} \\ \hline
    \endhead

    \hline \multicolumn{7}{r}{{Continued on next page}} \\ \hline
    \endfoot

    \hline
    \endlastfoot

    \tiny
    $z \geq 5$          &  0.31780 &  1.90456 &  3.49696 &  1.59685 & -0.81864 &  0.05797 \\ 
    $z \geq 5$          &  0.02818 &  0.48614 &  3.38461 & 11.48542 & 17.92994 & 10.52881 \\ 
    $4.5 \leq z < 5$    & -0.08044 & -0.15688 & -0.18515 & -0.95701 & -1.34354 &  0.01552 \\ 
    $4.5 \leq z < 5$    & -0.07063 & -1.13526 & -6.99940 &-20.95482 &-31.48083 &-18.81004 \\ 
    $4.0 \leq z < 4.5$  &  0.05670 &  0.53467 &  1.11348 &  0.18469 & -0.90595 &  0.01670 \\ 
    $4.0 \leq z < 4.5$  &  0.02399 &  0.48463 &  3.81766 & 14.28906 & 24.61804 & 16.15502 \\ 
    $3.5 \leq z < 4.0$  &  0.30591 &  2.06795 &  4.69183 &  3.98841 &  0.66544 &  0.01872 \\ 
    $3.5 \leq z < 4.0$  &  0.09243 &  1.86077 & 14.65700 & 56.10323 &103.60800 & 74.62733 \\ 
    $3.0 \leq z < 3.5$  & -0.63885 & -2.69953 & -3.68593 & -1.82274 & -0.54234 & -0.00207 \\ 
    $3.0 \leq z < 3.5$  & -0.00255 &  0.01824 &  0.61534 &  3.51845 &  6.81554 &  4.55758 \\ 
    $2.5 \leq z < 3.0$  & -0.97712 & -4.55106 & -7.37515 & -5.01543 & -1.57583 &  0.01368 \\ 
    $2.5 \leq z < 3.0$  &  0.01384 &  0.29867 &  2.43638 &  9.15021 & 15.19302 &  9.38886 \\ 
    $2.0 \leq z < 2.5$  & -0.77190 & -3.64542 & -5.87739 & -3.81529 & -1.17848 & -0.01827 \\ 
    $2.0 \leq z < 2.5$  &  0.01596 &  0.33515 &  2.66812 &  9.82456 & 16.11500 &  9.86593 \\ 
    $1.5 \leq z < 2.0$  &  0.07274 &  0.76975 &  2.64941 &  3.84594 &  2.01147 &  0.34789 \\ 
    $1.5 \leq z < 2.0$  &  0.00959 &  0.21865 &  1.87183 &  7.33157 & 12.60437 &  8.14784 \\ 
    $1.0 \leq z < 1.5$  &  0.03566 &  0.33147 &  1.16261 &  1.97365 &  1.23036 &  0.24592 \\ 
    $1.0 \leq z < 1.5$  &  0.00910 &  0.22525 &  2.09226 &  8.92938 & 16.99349 & 12.23198 \\ 
    $0.5 \leq z < 1.0$  &  0.00033 &  0.01233 &  0.14861 &  0.64626 &  0.58232 &  0.08427 \\ 
    $z < 0.5$           &  0.00012 &  0.00480 &  0.06104 &  0.26417 &  0.07170 & -0.03721 \\ 
\end{longtable}

\end{onecolumn}

For most corrections, a single polynomial fit was unable to capture accurately the shape of the correction, even when higher order polynomials were tested, so 2 fifth order polynomials are used for most corrections. Also, due to the low number of SMBH present at higher redshift, in the TNG100 box, the correction calculated at z = 5 is applied to all SMBH with a larger redshift. As shown in Figure \ref{fig:correction_figure}, the corrections at z = 4.5 and z = 5 are very similar, therefore, this compromise is acceptable. The coefficients of the polynomial fits are presented in Table \ref{tab:resolution}.

\end{appendix} 


\bsp	
\label{lastpage}
\end{document}